\documentclass[pra,aps,amsmath,amssymb,showkeys,showpacs]{revtex4}

\newcommand{\hh}{\mathcal{H}}
\newcommand{\lnp}{\mathcal{L}}

\newcommand{\lsp}{\mathcal{L}_{+}}
\newcommand{\bro}{\boldsymbol{\rho}}
\newcommand{\wbro}{\widetilde{\boldsymbol{\rho}}}
\newcommand{\wbom}{\widetilde{\boldsymbol{\omega}}}
\newcommand{\weta}{\widetilde{\boldsymbol{\eta}}}
\newcommand{\wsig}{\widetilde{\boldsymbol{\sigma}}}
\newcommand{\wvpi}{\boldsymbol{\varpi}}
\newcommand{\bsg}{\boldsymbol{\sigma}}
\newcommand{\pen}{\openone}
\newcommand{\bnil}{\mathbf{0}}
\newcommand{\am}{\mathsf{A}}
\newcommand{\ax}{\mathsf{X}}
\newcommand{\ay}{\mathsf{Y}}
\newcommand{\nm}{\mathsf{N}}
\newcommand{\tr}{\mathrm{tr}}
\newcommand{\clb}{\mathcal{B}}
\newcommand{\mc}{\mathcal{M}}
\newcommand{\nc}{\mathcal{N}}
\newcommand{\cmb}{\mathbb{B}}
\newcommand{\wrn}{\widetilde{R}}
\newcommand{\wrh}{\widetilde{H}}

\begin{document}
\clearpage
\preprint{}

\title{Uncertainty relations for MUBs and SIC-POVMs in terms of generalized entropies}
\author{Alexey E. Rastegin}
\affiliation{Department of Theoretical Physics, Irkutsk State University,
Gagarin Bv. 20, Irkutsk 664003, Russia}

\begin{abstract}
We formulate uncertainty relations for mutually
unbiased bases and symmetric informationally complete measurements
in terms of the R\'{e}nyi and Tsallis entropies. For arbitrary
number of mutually unbiased bases in a finite-dimensional Hilbert
space,  we give a family of Tsallis $\alpha$-entropic bounds for
$\alpha\in(0;2]$. Relations in a model of detection
inefficiences are obtained. In terms of R\'{e}nyi's entropies,
lower bounds are given for $\alpha\in[2;\infty)$. State-dependent
and state-independent forms of such bounds are both given.
Uncertainty relations in terms of the min-entropy are separately
considered. We also obtain lower bounds in term of the so-called
symmetrized entropies. The presented results for mutually unbiased
bases are extensions of some bounds previously derived
in the literature. We further formulate new properties of
symmetric informationally complete measurements in a
finite-dimensional Hilbert space. For a given state and any
SIC-POVM, the index of coincidence of generated probability
distribution is exactly calculated. Short notes are made on
potential use of this result in entanglement detection. Further,
we obtain state-dependent entropic uncertainty relations for a
single SIC-POVM. Entropic bounds are derived in terms of the
R\'{e}nyi $\alpha$-entropies for $\alpha\in[2;\infty)$ and the
Tsallis $\alpha$-entropies for $\alpha\in(0;2]$. In the Tsallis
formulation, a case of detection inefficiences is briefly
mentioned. For a pair of symmetric informationally complete
measurements, we also obtain an entropic bound of Maassen--Uffink
type.
\end{abstract}

\maketitle

\section{Introduction}\label{sec1}

Heisenberg's uncertainty principle \cite{heisenberg} and the
concept of complementarity are basic in understanding quantum
limitations. For two noncommuting observables, one cannot find a
basis of common eigenstates in a Hilbert space. Hence, we should
have some trade-off relations between two probability
distributions. There are many ways to express quantitatively such
relations \cite{lahti,ww10,brud11}. Due to recent advances in
quantum information properties, the uncertainty principle often
get a new elucidation \cite{shih03,dmg07}. Entropic functionals
provide a natural and flexible tool for expressing an uncertainty
in quantum measurements \cite{maass}. Entropic uncertainty
relations were studied in many cases (see reviews
\cite{ww10,brud11} and references therein). Together with the
Shannon entropy, other entropies are of interest. Utility of
entropic uncertainty relations with parametric dependence was
originally mentioned in reference \cite{maass}. Entropic trade-off
between more than two measurements are now the subject of active
research \cite{ww10,aamb10}. This issue is essential in own rights
as well as in studying the security of quantum cryptographic
protocols \cite{dmg07,ngbw12}. Entropic trade-off relations were
recently formulated for quantum operations \cite{rprz12}.
Uncertainty relations for mutually unbiased bases (MUBs) are
interesting in various respects \cite{molm09a,petz10}. Such bases
with an explicit construction for them were actually considered by
Schwinger \cite{schwinger}.

If two bases are mutually unbiased, then the overlaps between any
basis state in one basis and all basis states in the other are the
same. The detection of a particular basis state reveals no
information about the state, which was prepared in another basis.
This fact is essential in some popular schemes of quantum
cryptography. Mutually unbiased bases are used in quantum state
reconstruction \cite{wf89}, quantum error correction
\cite{gott96,cald97}, detection of quantum entanglement
\cite{shbah12}, and the mean king's problem \cite{vaid87,ena01}.
Many aspects of mutually unbiased bases are reviewed in reference
\cite{bz10}. Maximal sets of $d+1$ mutually unbiased bases have
been built for the case, when $d$ is power of a prime number. If
the dimensionality is another composite number, maximal sets of
MUBs are an open problem \cite{bz10}. Entropic uncertainty
relations for $d+1$ mutually unbiased bases in $d$-dimensional
Hilbert space were obtained in references \cite{ivan92,sanchez93}.
For even $d$, this relation has been improved \cite{jsan95}.
S\'{a}nchez \cite{sanchez93} also gave the exact bounds for the
qubit case $d=2$. These exact relations have been extended in
terms of Tsallis' entropies \cite{rastqip}. Specific uncertainty
relations for MUBs and some of their applications were given in
reference \cite{ballester}. These formulations are given in terms
of the Shannon entropy and the R\'{e}nyi entropy of order $2$.
They are based on the fact that $d+1$ MUBs, if exist, form a
spherical $2$-design in $d$-dimensional Hilbert space. This notion
is also connected with symmetric informationally complete
measurements (SIC-POVMs) \cite{rbsc04,rottler}.

In this work, we obtain uncertainty bounds in terms of the
R\'{e}nyi and Tsallis entropies for MUBs and for SIC-POVMs.
Studying statistical properties of such measurements are motivated
by applications of complementarity and uncertainty relations in
quantum information processing \cite{molm09a}. The use of
generalized entropies may allow to extend a potential scope of
obtained relations. As an example, we recall the formulation of
Bell's theorem in terms of the conditional Tsallis entropies
\cite{rastqqt}. It has allowed to extend a class of probabilistic
models, whose non-locality or contextuality are testable within
entropic approach. The paper is organized as follows. In Section
\ref{sec2}, the concepts of MUBs and SIC-POVMs are briefly
recalled. Definitions of the R\'{e}nyi and Tsallis entropies are
discussed as well. In Section \ref{sec3}, uncertainty relations
for an arbitrary number of MUBs are derived in terms of Tsallis'
$\alpha$-entropies for $\alpha\in(0;2]$ and R\'{e}nyi's
$\alpha$-entropies for $\alpha\in[2;\infty)$. For any set of MUBs,
uncertainty relations of Maassen--Uffink type is expressed by
means of the symmetrized entropies. In Section \ref{sec4}, we
examine entropic uncertainty relations for SIC-POVMs. For a single
SIC-POVM, state-independent lower bounds on its Tsallis entropy of
order $\alpha\in(0;2]$ and on its R\'{e}nyi's entropy of order
$\alpha\in[2;\infty)$ are derived. For a pair of SIC-POVMs,
entropic bounds of Maassen--Uffink type are considered. In this
regard, we give a new short proof of Tsallis-entropy and
Renyi-entropy uncertainty relations for two rank-one POVMs. In
Section \ref{sec5}, we conclude the paper with a summary of
results.

\section{Definitions and notation}\label{sec2}

In this section, the required material is reviewed. First, we
recall the notions of mutually unbiased bases and symmetric
informationally complete POVMs. Second, we briefly consider the
R\'{e}nyi and Tsallis entropies.

\subsection{Mutually unbiased bases and symmetric informationally complete POVMs}\label{sc21}

Let $\lnp(\hh)$ be the space of linear operators on
$d$-dimensional Hilbert space $\hh$. By $\lsp(\hh)$, we denote the
set of positive semidefinite operators on $\hh$. A density
operator $\bro\in\lsp(\hh)$ has unit trace, i.e. $\tr(\bro)=1$.
For $\ax,\ay\in{\mathcal{L}}(\hh)$, the Hilbert--Schmidt inner
product is defined as \cite{watrous1}
\begin{equation}
\langle\ax{\,},\ay\rangle_{\rm{hs}}:=\tr(\ax^{\dagger}\ay)
\ . \label{hsdef}
\end{equation}
Let $\clb_{1}=\bigl\{|b_{j}^{(1)}\rangle\bigr\}$ and
$\clb_{2}=\bigl\{|b_{k}^{(2)}\rangle\bigr\}$ be two orthonormal
bases in a $d$-dimensional Hilbert space $\hh$. They are
said to be mutually unbiased if and only if for all $j$ and $k$,
\begin{equation}
\bigl|\langle{b}_{j}^{(1)}|b_{k}^{(2)}\rangle\bigr|=\frac{1}{\sqrt{d}}
\ . \label{twb}
\end{equation}
The set $\cmb=\bigl\{\clb_{1},\ldots,\clb_{M}\bigr\}$ is a set of
mutually unbiased bases, when each two bases from this set are
mutually unbiased. The states within MUBs are indistinguishable in
the following sense. If the two observables have unbiased
eigenbases, then the measurement of one observable reveals no
information about possible outcomes of the measurement of other.
This property is used in the BB84 protocol \cite{bb84} and the
six-state protocol \cite{bruss98}. In two dimensions, three
eigenbases of the Pauli observables are mutually unbiased. Basic
constructions for MUBs have been proposed in references
\cite{wf89,bbrv02,kr04,wb05} and reviewed in reference
\cite{bz10}. A search of MUBs in dimensions, which are not a prime
power, is closely related to studies of complex Hadamard matrices
\cite{bz07}.

In general, quantum measurements share many subtle properties,
which are important in quantum information \cite{peresq}. In the
presence of additive conserved quantities, repeatable measurements
turn out to be limited according to the Wigner--Araki--Yanase
theorem \cite{loveb11}. Generalized quantum measurements are
commonly treated within the POVM formalism \cite{peresq}. A
positive operator-valued measure $\nc=\{\nm_{j}\}$ is a set of
elements $\nm_{j}\in\lsp(\hh)$ obeying the completeness relation
\begin{equation}
\sum\nolimits_{j} \nm_{j}=\pen
\ . \label{cmprl}
\end{equation}
Here, the symbol $\pen$ denotes the identity operator on $\hh$.
It is of key importance that a number of different
outcomes can be more than the dimensionality of $\hh$. For
pre-measurement state $\bro$, the probability of $j$th outcome is
written as \cite{peresq}
\begin{equation}
p_{j}(\nc|\bro)=\tr(\nm_{j}\bro)
\ . \label{njpr}
\end{equation}
POVM measurements are an indispensable tool in quantum information
theory \cite{watrous1}. A POVM is said to be informationally
complete, if its statistics determine completely the quantum state
\cite{prug77,busch91,dps04}. In $d$-dimensional space, we take a
set of $d^{2}$ rank-one operators of the form
$\nm_{j}=d^{-1}|\phi_{j}\rangle\langle\phi_{j}|$. When the
normalized vectors $|\phi_{j}\rangle$ obey
\begin{equation}
\bigl|\langle\phi_{j}|\phi_{k}\rangle\bigr|^{2}=\frac{1}{d+1}
\ , \qquad j\neq{k}
\ , \label{undn1}
\end{equation}
the set $\{\nm_{j}\}$ is called a symmetric informationally
complete POVM \cite{rbsc04}. Weyl--Heisenberg (WH) covariant
SIC-sets of states in prime dimensions are examined in reference
\cite{adf07}. WH SIC-sets, whenever they exist, consist solely of
minimum uncertainty states with respect to R\'{e}nyi's $2$-entropy
for a complete set of MUBs \cite{adf07}. Klappenecker et al.
\cite{krsw05} discussed approximate versions of SIC-POVMs, where a
small deviation from uniformity of the inner products is allowed.
It follows from equation (\ref{undn1}) that pairwise inner
products between the POVM elements are all
\begin{equation}
\langle\nm_{j}{\,},\nm_{k}\rangle_{\rm{hs}}=\frac{1}{d^{2}(d+1)}
\ , \qquad j\neq{k}
\ . \label{sicdf}
\end{equation}
From equations (\ref{undn1}) and (\ref{sicdf}), basic properties
of a SIC-POVM can be derived \cite{rbsc04}. In particular, the
vectors $|\phi_{j}\rangle$ form a spherical $2$-design
\cite{rbsc04,rottler}. This consequence has been derived in
reference \cite{rottler}. Here, we refrain from considering
spherical designs. Instead, we will present an exact calculation
of the index of coincidence for probability distribution generated
by a SIC-POVM.

\subsection{Renyi's and Tsallis' entropies}\label{sc22}

The R\'{e}nyi and Tsallis entropies both form an especially
important family of one-parametric generalizations of the Shannon
entropy. For $\alpha>0\neq1$, the R\'{e}nyi $\alpha$-entropy is
defined as \cite{renyi61}
\begin{equation}
R_{\alpha}(p):=\frac{1}{1-\alpha}{\>}{\ln}{\left(\sum\nolimits_{j} p_{j}^{\alpha}
\right)}
{\,}. \label{renent}
\end{equation}
In the limit $\alpha\to1$, this expression leads to the standard
Shannon entropy $H_{1}(p)=-\sum_{j}p_{j}\ln{p}_{j}$. The entropy
(\ref{renent}) is a non-increasing function of the order $\alpha$
\cite{renyi61}. Basic properties of the R\'{e}nyi entropy and some of
its physical applications are reviewed in reference \cite{ja04}.
Together with the standard case $\alpha=1$, the following two
choices of $\alpha$ are particularly important. For $\alpha=2$,
formula (\ref{renent}) gives the so-called collision entropy
\begin{equation}
R_{2}(p)=-\ln\left(\sum\nolimits_{j}p_{j}^{2}\right)
{\,}. \label{clen}
\end{equation}
The collision-entropy uncertainty relations for MUBs were obtained
in references \cite{ballester,molm09}. In the limit
$\alpha\to\infty$, we further obtain the min-entropy
\begin{equation}
R_{\infty}(p)=-\ln\bigl(\max{p}_{j}\bigr)
{\>}. \label{mnen}
\end{equation}
The min-entropy is of particular interest in cryptography
\cite{ngbw12}, and is also related to the extrema of the discrete
Wigner function \cite{MWB10}. For the dimensionality $d=2^{n}$,
the writers of reference \cite{MWB10} derived lower bounds on the
sum of min-entropies for several MUBs.

For $\alpha>0\neq1$, the Tsallis $\alpha$-entropy of probability
distribution $\{p_{j}\}$ is defined by \cite{tsallis}
\begin{equation}
H_{\alpha}(p):=\frac{1}{1-\alpha}{\,}\left(\sum\nolimits_{j} p_{j}^{\alpha}
- 1 \right)
{\,}. \label{tsaent}
\end{equation}
With slightly other factor, the functional (\ref{tsaent}) was
previously derived within a formal context by Havrda and
Charv\'{a}t \cite{HC67}. In statistical physics, the entropy
(\ref{tsaent}) was introduced and motivated by Tsallis
\cite{tsallis}. Basic properties of the entropy (\ref{tsaent}) and
its conditional form are discussed in references
\cite{sf06,rastkyb}. It is useful to rewrite equation
(\ref{tsaent}) in terms of the $\alpha$-logarithm, which is a
standard tool in non-extensive thermostatistics. For
$\alpha>0\neq1$ and $x>0$, the $\alpha$-logarithm is defined as
\begin{equation}
\ln_{\alpha}(x):=\frac{x^{1-{\alpha}}-1}{1-{\alpha}}
\ . \label{aldf}
\end{equation}
The right-hand side of equation (\ref{tsaent}) can be represented
in two equivalent ways, namely
\begin{equation}
H_{\alpha}(p)=-\sum\nolimits_{j}p_{j}^{\alpha}{\,}\ln_{\alpha}(p_{j})
=\sum\nolimits_{j}p_{j}{\>}{\ln_{\alpha}}{\left(\frac{1}{p_{j}}\right)}
{\,}. \label{tsaent2}
\end{equation}
Taking $\alpha\to1$, the $\alpha$-logarithm is reduced to the
standard logarithm $\ln{x}$. Here, the entropy (\ref{tsaent})
recovers the Shannon entropy $H_{1}(p)=-\sum_{j}p_{j}\ln{p}_{j}$.
The Tsallis entropy (\ref{tsaent}) is a concave function of its
entry. Namely, for all $\lambda\in[0;1]$ and two probability
distributions $p=\{p_{j}\}$ and $q=\{q_{j}\}$, we have
\begin{equation}
{H_{\alpha}}{\bigl(\lambda{p}+(1-\lambda)q\bigr)}\geq
\lambda{\,}H_{\alpha}(p)+(1-\lambda)H_{\alpha}(q)
\ . \label{hacn}
\end{equation}
Indeed, the function $x\mapsto\bigl(x^{\alpha}-x\bigr)/(1-\alpha)$
is concave. It must be stressed that the R\'enyi $\alpha$-entropy
does not obey the concavity property for $\alpha>1$. Namely, it is
not purely convex nor purely concave for such $\alpha$
\cite{ja04}.

In a case of detection inefficiencies, we will use the following
item. To the given value $\eta\in[0;1]$ and probability
distribution $\{p_{j}\}$, we assign a ``distorted'' distribution:
\begin{equation}
p_{j}^{(\eta)}=\eta{\,}p_{j}
\ , \qquad
p_{\varnothing}^{(\eta)}=1-\eta
\ . \label{dspd}
\end{equation}
The probability $p_{\varnothing}^{(\eta)}$ will correspond to the
no-click event. As was shown in reference \cite{rastqqt}, for all
$\alpha>0$ we have
\begin{equation}
{H_{\alpha}}{\bigl(p^{(\eta)}\bigr)}=\eta^{\alpha}H_{\alpha}(p)+h_{\alpha}(\eta)
\ . \label{qtlm0}
\end{equation}
Here, the binary Tsallis entropy $h_{\alpha}(\eta)$ is expressed by
\begin{equation}
h_{\alpha}(\eta)=-\eta^{\alpha}\ln_{\alpha}(\eta)-(1-\eta)^{\alpha}\ln_{\alpha}(1-\eta)
\ . \label{bnta}
\end{equation}
Results of such a kind have been used in studying entropic Bell
inequalities with detector inefficiencies \cite{rchtf12}.
Applications of the R\'{e}nyi and Tsallis entropies in quantum
information theory are discussed in book \cite{bengtsson}.
Conditional R\'{e}nyi and Tsallis entropies of partitions on
quantum logic are examined in reference \cite{rastctp}.

\section{Lower entropic bounds for MUBs}\label{sec3}

In this section, we formulate entropic uncertainty relations for
mutually unbiased bases. First, we derive entropic lower bounds
for an arbitrary number of MUBs in terms of the Tsallis
$\alpha$-entropies of order $\alpha\in(0;2]$. Second, the
R\'{e}nyi formulation is considered. Finally, we present entropic
bounds of Maassen--Uffink type in terms of the so-called
symmetrized entropies.

\subsection{Uncertainty relations for arbitrary set of MUBs}\label{ssc31}

We begin with uncertainty relations in terms of Tsallis'
$\alpha$-entropies for $\alpha\in(0;2]$. Since state-dependent and
state-independent formulations are both of interest, we discuss
them below. The following statement takes place.

\newtheorem{prop1}{Proposition}
\begin{prop1}\label{pan1}
Let $\cmb=\bigl\{\clb_{1},\ldots,\clb_{M}\bigr\}$ be a set of MUBs
in $d$-dimensional Hilbert space $\hh$. For $\alpha\in(0;2]$ and
arbitrary density matrix $\bro$ on $\hh$, the sum of Tsallis'
entropies satisfies the state-dependent bound
\begin{equation}
\frac{1}{M}\sum_{\clb\in\cmb} H_{\alpha}(\clb|\bro)
\geq{\ln_{\alpha}}{\left(\frac{Md}{\tr(\bro^{2}){\,}d+M-1}\right)}
{\,}. \label{fmub0}
\end{equation}
The state-independent formulation is expressed as
\begin{equation}
\frac{1}{M}\sum_{\clb\in\cmb} H_{\alpha}(\clb|\bro)
\geq{\ln_{\alpha}}{\left(\frac{Md}{d+M-1}\right)}
{\,}. \label{fmub1}
\end{equation}
\end{prop1}

{\bf Proof.} First, we consider the case $\alpha\neq1$. To each
mutually unbiased basis $\clb=\bigl\{|b_{j}\rangle\bigr\}$, we
assign the so-called index of coincidence
\begin{equation}
C(\clb|\bro):=\sum\nolimits_{j=1}^{d}p_{j}(\clb|\bro)^{2}
\ , \label{indf}
\end{equation}
where $p_{j}(\clb|\bro)=\langle{b}_{j}|\bro|b_{j}\rangle$.
The second derivative of function $\ln_{\alpha}(1/x)$ is equal to
$(2-\alpha){\,}x^{\alpha-3}$ and positive for $\alpha\leq2$.
Hence, the function itself is convex. Combining Jensen's
inequality with the right-hand side of equation (\ref{tsaent2}),
for $0<\alpha\leq2$ and any probability distribution we have
\begin{equation}
H_{\alpha}(p)\geq{\ln_{\alpha}}{\left(\frac{1}{C(p)}\right)}
{\,}. \label{conc1}
\end{equation}
Using this inequality and Jensen's inequality again, one gives
\begin{equation}
\frac{1}{M}\sum_{m=1}^{M} H_{\alpha}(\clb_{m}|\bro)\geq
\sum_{m=1}^{M}\frac{1}{M}{\>}{\ln_{\alpha}}{\left(\frac{1}{C_{m}}\right)}
\geq{\ln_{\alpha}}{\left\{\left(\frac{1}{M}\sum\nolimits_{m}C_{m}\right)^{-1}\right\}}
{\,}, \label{conc2}
\end{equation}
where $C_{m}=C(\clb_{m}|\bro)$. The writers of reference
\cite{molm09} have shown the following. For $M$ mutually unbiased
bases, the indices of coincidence obey
\begin{equation}
\sum\nolimits_{m=1}^{M}C_{m}\leq\tr(\bro^{2})+\frac{M-1}{d}
\ . \label{lwbm}
\end{equation}
As the function $x\mapsto\ln_{\alpha}(1/x)$ decreases, the
inequality $x\leq{y}$ implies
$\ln_{\alpha}(1/x)\geq\ln_{\alpha}(1/y)$. Combining this with
equations (\ref{conc2}) and (\ref{lwbm}) finally gives the claim
(\ref{fmub0}). Substituting $\tr(\bro^{2})\leq1$ into equation
(\ref{lwbm}), the same reason provides the bound (\ref{fmub1}).
Finally, the standard case $\alpha=1$ is obtained in the limit
$\alpha\to1$.
$\blacksquare$

For pure states, we have $\tr(\bro^{2})=1$. Then the
state-dependent bound (\ref{fmub0}) is reduced to the form
(\ref{fmub1}). For impure states, the lower bound (\ref{fmub0}) is
stronger due to $\tr(\bro^{2})<1$. That is, obtained lower bounds
increase with a deviation from purity. In the limit $\alpha\to1$,
the bounds (\ref{fmub0}) and (\ref{fmub1}) reproduce the relations
of reference \cite{molm09}. Unlike previous results, the entropic
bounds of reference \cite{molm09} hold for any dimension $d$ and
any number $M$ of MUBs. The only assumption is that these MUBs
merely exist. Thus, we have extended general entropic inequalities
for MUBs to one-parametric family of $\alpha$-entropies for all
$\alpha\in(0;2]$. Note that our reasons differ from the method of
reference \cite{molm09}. The latter is mainly based on the results
of reference \cite{ht01}. Its authors gave inequalities between
the Shannon entropy and the index of coincidence \cite{ht01}. Such
inequalities have also been used to obtain more detailed bounds of
state-independent form \cite{molm09}. As was exemplified therein,
entropic bounds of such a kind are not stronger than equation
(\ref{fmub0}) for $\alpha=1$. Nevertheless, it would be
interesting to develop this issue for the Tsallis
$\alpha$-entropies. Taking $M=d+1$, the inequalities (\ref{fmub0})
and (\ref{fmub1}) give
\begin{align}
\frac{1}{d+1}{\,}\sum\nolimits_{m=1}^{d+1} H_{\alpha}(\clb_{m}|\bro)
&\geq{\ln_{\alpha}}{\left(\frac{d+1}{\tr(\bro^{2})+1}\right)}
\label{fmubd0}\\
&\geq{\ln_{\alpha}}{\left(\frac{d+1}{2}\right)}
{\,}. \label{fmubd1}
\end{align}
When $\alpha=1$, the right-hand side of equation
(\ref{fmubd1}) leads to the well-known entropic bound.
It has been given by S\'{a}nchez \cite{sanchez93} and
later rederived in many works. In particular, the following
fact is suitable here \cite{ballester}. Any union of $d+1$ MUBs in
$d$-dimensional Hilbert space forms a spherical $2$-design
\cite{rottler}. Our bound (\ref{fmubd1}) could be derived
from this fact and equation (\ref{conc1}). However, sets
of $d+1$ MUBs have been built only in Hilbert spaces of a
prime-power dimensionality. So, it is of importance to have
entropic bounds for arbitrarily taken number of MUBs.

Entropic uncertainty bounds with a parametric dependence are not
only a way to write the principle itself \cite{maass}. Entropic
uncertainty relations impose some conditions on measurement
probabilities. Thus, the relations (\ref{fmub0}) and (\ref{fmub1})
may be of practical interest. In this regard, we also consider a
more realistic case of detection inefficiencies. Let the parameter
$\eta\in[0;1]$ characterize a detector efficiency. The no-click
event is presented by additional outcome $\varnothing$. Assume
that for any base $\clb\in\cmb$ the inefficiency-free distribution
is distorted according to equation (\ref{dspd}):
\begin{equation}
p_{j}^{(\eta)}(\clb|\bro)=\eta{\,}p_{j}(\clb|\bro)
\ , \qquad
p_{\varnothing}^{(\eta)}(\clb|\bro)=1-\eta
\ . \label{dspd1}
\end{equation}
In other words, we mean detectors of the same efficiency for all
treated MUBs. From the physical viewpoint, this assumption is
quite natural. Using equations (\ref{qtlm0}) and (\ref{fmub0}),
for $\alpha\in(0;2]$ we obtain
\begin{equation}
\frac{1}{M}\sum_{\clb\in\cmb} H_{\alpha}^{(\eta)}(\clb|\bro)
\geq\eta^{\alpha}{\,}{\ln_{\alpha}}{\left(\frac{Md}{\tr(\bro^{2}){\,}d+M-1}\right)}
+h_{\alpha}(\eta)
{\>}. \label{fmub0et}
\end{equation}
Here, the $H_{\alpha}^{(\eta)}(\clb|\bro)$ denotes the entropy of
the distribution (\ref{dspd1}). The result (\ref{fmub0et}) is an
entropic uncertainty relation in the model of detection
inefficiencies. In the case $\alpha=1$, the inefficiency-free
lower bound is merely added by the binary Shannon entropy
$h_{1}(\eta)$. Entropies of actual probability
distributions take into account not only quantum uncertainties. An
additional uncertainty is inevitably inserted by the detector. The
right-hand side of equation (\ref{fmub0et}) also allows to
estimate a required amount of detector efficiency. The first term
therein should be sufficiently large in comparison with the binary
entropy $h_{\alpha}(\eta)$.

\newtheorem{prop2}[prop1]{Proposition}
\begin{prop2}\label{pan2}
Let $\cmb=\bigl\{\clb_{1},\ldots,\clb_{M}\bigr\}$ be a set of MUBs
in $d$-dimensional Hilbert space $\hh$. For $\alpha\in[2;\infty)$
and arbitrary density matrix $\bro$ on $\hh$, the sum of
R\'{e}nyi's entropies satisfies the state-dependent bound
\begin{equation}
\frac{1}{M}\sum_{\clb\in\cmb} R_{\alpha}(\clb|\bro)\geq
\frac{\alpha}{2(\alpha-1)}{\>}{\ln}{\left(\frac{Md}{\tr(\bro^{2}){\,}d+M-1}\right)}
{\,}. \label{rmub0}
\end{equation}
The state-independent formulation is expressed as
\begin{equation}
\frac{1}{M}\sum_{\clb\in\cmb} R_{\alpha}(\clb|\bro)\geq
\frac{\alpha}{2(\alpha-1)}{\>}{\ln}{\left(\frac{Md}{d+M-1}\right)}
{\,}. \label{rmub1}
\end{equation}
\end{prop2}

{\bf Proof.} For $\alpha\geq2$, we have the inequality
\begin{equation}
\left(\sum\nolimits_{j}p_{j}^{\alpha}\right)^{1/\alpha}
\leq\left(\sum\nolimits_{j}p_{j}^{2}\right)^{1/2}=C(p)^{1/2}
\ . \label{nmq2}
\end{equation}
This relation follows from theorem 19 of book \cite{hardy}. The
function $x\mapsto(1-\alpha)^{-1}\ln{x}$ decreases for $\alpha>1$.
Combining this with
$\sum\nolimits_{j}p_{j}^{\alpha}\leq{C}^{\alpha/2}$, for
$\alpha\geq2$ and any probability distribution we get
\begin{equation}
R_{\alpha}(p)\geq\frac{\alpha}{2(1-\alpha)}{\>}\ln{C}(p)
\ . \label{rain}
\end{equation}
Further, the function $x\mapsto(1-\alpha)^{-1}\ln{x}$ is convex
for $\alpha>1$, whence
\begin{equation}
\frac{1}{M}\sum_{m=1}^{M} R_{\alpha}(\clb_{m}|\bro)\geq
\sum_{m=1}^{M}\frac{1}{M}{\>}\frac{\alpha}{2(1-\alpha)}{\>}\ln{C}_{m}
\geq\frac{\alpha}{2(1-\alpha)}{\>}{\ln}{\left(\frac{1}{M}\sum\nolimits_{m}C_{m}\right)}
{\,}. \label{concr2}
\end{equation}
As the function $x\mapsto(1-\alpha)^{-1}\ln{x}$ is decreasing for
such $\alpha$, the inequalities (\ref{lwbm}) and (\ref{concr2})
provide the claim (\ref{rmub0}). The inequality (\ref{rmub1}) is
obtained by substituting $\tr(\bro^{2})\leq1$ into equation
(\ref{rmub0}).
$\blacksquare$

With respect to a dependence on density matrix $\bro$, the lower
bound (\ref{rmub0}) is similar to equation (\ref{fmub0}). It is
reduced to the state-independent form (\ref{rmub1}) for all pure
states. For impure states, the bound (\ref{rmub0}) is strictly
stronger than equation (\ref{rmub1}). For $\alpha\in[2;\infty)$,
we obtained the lower bounds (\ref{rmub0}) and (\ref{rmub1}),
which both depend on $\alpha$. Substituting $\alpha=2$, these
inequalities leads to
\begin{align}
\frac{1}{M}\sum_{\clb\in\cmb} R_{2}(\clb|\bro)
&\geq
\ln\!\left(\frac{Md}{\tr(\bro^{2}){\,}d+M-1}\right)
\label{rmub0012}\\
&\geq\ln\!\left(\frac{Md}{d+M-1}\right)
{\,}. \label{rmub012}
\end{align}
This result was given in reference \cite{molm09}. Thus, we have
obtained its one-parametric extension. When $d$ is a prime and
$M=d+1$, the inequality (\ref{rmub012}) is saturated with every WH
fiducial vector \cite{adf07}. Since the R\'{e}nyi $\alpha$-entropy
does not increase with $\alpha$, the bound (\ref{rmub0012}) holds
for all Renyi's entropies of order $\alpha\in(0;2)$. For such
orders, our bound is independent of $\alpha$. In the limit
$\alpha\to\infty$, we have a lower bound on the sum of
min-entropies:
\begin{align}
\frac{1}{M}\sum_{\clb\in\cmb} R_{\infty}(\clb|\bro)
&\geq
\frac{1}{2}{\>}{\ln}{\left(\frac{Md}{\tr(\bro^{2}){\,}d+M-1}\right)}
\label{rmub00in}\\
&\geq\frac{1}{2}{\>}{\ln}{\left(\frac{Md}{d+M-1}\right)}
{\,}. \label{rmub0in}
\end{align}
For dimensionality $d=2^{n}$, min-entropy lower bounds of
state-independent form for arbitrary number of MUBs were derived
in reference \cite{MWB10}. In general, these bounds are stronger
than the right-hand side of equation (\ref{rmub0in}). We now
obtain an improvement of the bounds (\ref{rmub00in}) and
(\ref{rmub0in}). It is based on the inequality of Appendix
\ref{astt}.

\newtheorem{prop22}[prop1]{Proposition}
\begin{prop22}\label{pan22}
Let $\cmb=\bigl\{\clb_{1},\ldots,\clb_{M}\bigr\}$ be a set of MUBs
in $d$-dimensional Hilbert space $\hh$. For arbitrary density
matrix $\bro$ on $\hh$, the sum of min-entropies satisfies the
state-dependent bound
\begin{equation}
\frac{1}{M}\sum_{\clb\in\cmb} R_{\infty}(\clb|\bro)\geq
\ln{d}-
{\ln}{\left(1+M^{-1/2}\sqrt{d-1}\sqrt{\tr(\bro^{2}){\,}d-1}\right)}
{\,}. \label{rmub02}
\end{equation}
The state-independent formulation is expressed as
\begin{equation}
\frac{1}{M}\sum_{\clb\in\cmb} R_{\infty}(\clb|\bro)\geq
{\ln}{\left(\frac{\sqrt{M}{\,}d}{d+\sqrt{M}-1}\right)}
{\,}. \label{rmub12}
\end{equation}
\end{prop22}

{\bf Proof.} The function $x\mapsto-\ln{x}$ is convex. Combining
the corresponding Jensen inequality with equation (\ref{mnen})
gives
\begin{equation}
\frac{1}{M}\sum\nolimits_{m=1}^{M} R_{\infty}(\clb_{m}|\bro)\geq
{-\ln}{\left(\frac{1}{M}{\,}\sum\nolimits_{m}
\underset{j}{\max}{\,}p_{j}(\clb_{m}|\bro)\right)}
{\,}. \label{mrp1}
\end{equation}
The function $g_{d}(x)=d^{-1}\bigl(1+\sqrt{d-1}\sqrt{xd-1}\bigr)$
will be used for brevity. By the results (\ref{lwbm}) and
(\ref{aust}), we write
\begin{equation}
\frac{1}{M}\sum\nolimits_{m}
\underset{j}{\max}{\,}p_{j}(\clb_{m}|\bro)
\leq\frac{1}{M}\sum\nolimits_{m}g_{d}(C_{m})
\leq{g}_{d}{\left(\frac{1}{M}{\,}\sum\nolimits_{m}C_{m}\right)}
\leq{g}_{d}{\left(\frac{\tr(\bro^{2}){\,}d+M-1}{Md}\right)}
{\>}. \label{mrp2}
\end{equation}
Here, we used that the function $g_{d}(x)$ is concave and
increasing. The right-hand side of equation (\ref{mrp2}) can be
rewritten as
\begin{equation}
\frac{1}{d}\left(1+\sqrt{d-1}{\,}\sqrt{\frac{\tr(\bro^{2}){\,}d-1}{M}}{\,}\right)
\leq\frac{d+\sqrt{M}-1}{\sqrt{M}{\,}d}
{\>}, \label{mrp3}
\end{equation}
due to $\tr(\bro^{2})\leq1$. As the function $x\mapsto-\ln{x}$
decreases, the claims (\ref{rmub02}) and (\ref{rmub12})
immediately follow from equations (\ref{mrp1}) and (\ref{mrp3}).
$\blacksquare$

Up to a notation, the state-independent relation (\ref{rmub12})
coincides with equations (10) and (41) of reference \cite{MWB10}.
Thus, the min-entropy bound (\ref{rmub12}) is a state-dependent
extension of one of the main results of that paper. In addition,
our derivation does not assume a special choice of dimensionality
$d$. The only assumption is that $M$ mutually unbiased bases exist
for this dimensionality. For these reasons, the min-entropy lower
bound (\ref{rmub02}) is interesting itself as well as for
potential applications.

\subsection{Relations in terms of symmetrized entropies}\label{ssc32}

Most formulations of entropic uncertainty relations are of
Maassen--Uffink type \cite{maass} and pertain to a pair of
observables. Using such results, a lower bound on the sum of
Shannon entropies is easily given for arbitrary number of MUBs.
However, this bound is essentially weaker than the bounds obtained
in references \cite{ivan92,sanchez93,molm09}. Extensions of the
Maassen--Uffink uncertainty relations with quantum side
information have been obtained \cite{BCCRR10,ccyz12}. The
Maassen--Uffink approach \cite{maass} has been developed with use
of both the R\'{e}nyi \cite{birula06,rast10r} and Tsallis
entropies \cite{rast104,rast12num}, including the case of
quasi-Hermitian operators \cite{rast12quasi}. Here, we deal with
two generalized entropies, whose different orders are constrained
by a certain condition. This is predetermined by use of the Riesz
theorem. Therefore, for several measurements we should symmetrize
such entropies. For a pair of observables, relations in terms of
symmetrized entropies were derived within both the R\'{e}nyi
\cite{birula06} and Tsallis formulations \cite{raja95}. We will
now develop this issue for several MUBs. It is of interest,
because of the relations (\ref{fmub1}) and (\ref{rmub1}) deal with
the sums of Tsallis' and R\'{e}nyi's entropies of the same order.
For convenience, we recall a version of Riesz's theorem. It is a
statement about vector norms
\begin{equation}
\|v\|_{b}=\left(\sum\nolimits_{i} |v_{i}|^{b}\right)^{1/b}
, \label{nbvdf}
\end{equation}
where $b\geq1$. For positive vectors, we will also use
norm-like functionals $\|p\|_{\beta}=\left(\sum\nolimits_{j}
p_{j}^{\beta}\right)^{1/\beta}$ with $\beta>0$. Let us consider
two finite tuples of numbers $v\in{\mathbb{C}}^{m}$ and
$u\in{\mathbb{C}}^{n}$, which are connected by linear
transformation
\begin{equation}
v_{i}=\sum\nolimits_{j=1}^{n} t_{ij}{\,}u_{j}
\qquad (i=1,\ldots,m)
\ . \label{lntrns}
\end{equation}
Denoting
$\eta:={\max}\bigl\{|t_{ij}|:{\,}1\leq{i}\leq{m},{\,}1\leq{j}\leq{n}\bigr\}$,
a version of Riesz's theorem is posed as follows (see theorem 297
in book \cite{hardy}).

\newtheorem{lm1}{Lemma}
\begin{lm1}\label{lem1}
Let $a$ and $b$ be positive numbers such that $1/a+1/b=1$ and
$1<b<2$. If for all $u\in{\mathbb{C}}^{n}$ the matrix $[[t_{ij}]]$
satisfies
\begin{equation}
\|v\|_{2}\leq\|u\|_{2}
\ , \label{prcp}
\end{equation}
then the corresponding norms satisfy
\begin{equation}
\|v\|_{a}\leq{\eta}^{(2-b)/b}{\,}\|u\|_{b}
\ . \label{spps}
\end{equation}
\end{lm1}

The Riesz theorem is a relation between two vector norms with
conjugate indices, which obey $1/a+1/b=1$. In terms of the indices
$a$ and $b$, the entropic orders $\alpha$ and $\beta$ are put as
$\alpha=a/2$ and $\beta=b/2$ \cite{rast10r,rast104}. These orders
are constrained as $1/\alpha+1/\beta=2$. Assuming $s\in[0;1)$, we
can parametrize the entropic orders as
\begin{equation}
\max\{\alpha,\beta\}=\frac{1}{1-s}
\ , \qquad
\min\{\alpha,\beta\}=\frac{1}{1+s}
\ . \label{prms}
\end{equation}
so that $1/\alpha+1/\beta=2$. To write entropic uncertainty
relations for several MUBs, we shall use the symmetrized entropies
defined as
\begin{align}
\wrn_{s}(p)&:=\frac{1}{2}{\,}\bigl(R_{\alpha}(p)+R_{\beta}(p)\bigr)
\ , \label{rsm}\\
\wrh_{s}(p)&:=\frac{1}{2}{\,}\bigl(H_{\alpha}(p)+H_{\beta}(p)\bigr)
\ . \label{hsm}
\end{align}
With such entropies, uncertainty relations for a set of several
MUBs are expressed as follows.

\newtheorem{prop3}[prop1]{Proposition}
\begin{prop3}\label{pan3}
Let $\cmb=\bigl\{\clb_{1},\ldots,\clb_{M}\bigr\}$ be a set of MUBs
in $d$-dimensional Hilbert space $\hh$, and let $\bro$ be a
density matrix. For all $s\in[0;1)$, the symmetrized Tsallis
entropies satisfy
\begin{equation}
\frac{1}{M}\sum_{\clb\in\cmb} \wrh_{s}(\clb|\bro)\geq\frac{1}{2}{\>}\ln_{\mu}(d)
\ , \label{hsmub}
\end{equation}
where $\mu=(1-s)^{-1}$. For all $s\in[0;1)$, the symmetrized
R\'{e}nyi entropies satisfy
\begin{equation}
\frac{1}{M}\sum_{\clb\in\cmb} \wrn_{s}(\clb|\bro)\geq\frac{1}{2}{\>}\ln{d}
\ . \label{rsmub}
\end{equation}
\end{prop3}

{\bf Proof.} For
$\clb_{m}=\bigl\{|b_{i}^{(m)}\rangle\bigr\}$ and
$\clb_{n}=\bigl\{|b_{j}^{(n)}\rangle\bigr\}$, we have the
probabilities $q_{i}=\langle{b}_{i}^{(m)}|\bro|b_{i}^{(m)}\rangle$
and $p_{j}=\langle{b}_{j}^{(n)}|\bro|b_{j}^{(n)}\rangle$. Using
the transformation with matrix elements
\begin{equation}
t_{ij}=\langle{b}_{i}^{(m)}|b_{j}^{(n)}\rangle
\ , \qquad |t_{ij}|=d^{-1/2}
\ , \label{tbmn}
\end{equation}
Lemma \ref{lem1} leads to the relation
$\|q\|_{\alpha}\leq{d}^{(\beta-1)/\beta}\|p\|_{\beta}$ under the
conditions $1/\alpha+1/\beta=2$ and $1/2<\beta<1$. It is a special
case of the general result posed as lemma 2 of reference
\cite{rast104} (see also Proposition \ref{pan6} of the present
paper). Due to this relation between norm-like functions of the
probability distributions, the corresponding entropies satisfy
(cf. example 1 of reference \cite{rast104})
\begin{equation}
H_{\alpha}(\clb_{m}|\bro)+H_{\beta}(\clb_{n}|\bro)\geq\ln_{\mu}(d)
\ . \label{hhab}
\end{equation}
Swapping entropic orders in equation (\ref{hhab}), we get another
inequality for Tsallis entropies. A half-sum of these two
inequalities is represented as
\begin{equation}
\wrh_{s}(\clb_{m}|\bro)+\wrh_{s}(\clb_{n}|\bro)\geq\ln_{\mu}(d)
\qquad (m\neq{n})
\ . \label{symh2}
\end{equation}
From the given $M$ MUBs, we can choose $M(M-1)/2$ different pairs;
each MUB is appeared in $M-1$ pairs. Summing equation (\ref{symh2})
with respect to all pairs, we then obtain
\begin{equation}
(M-1){\,}\sum\nolimits_{m=1}^{M} \wrh_{s}(\clb_{m}|\bro)\geq\frac{M(M-1)}{2}{\>}\ln_{\mu}(d)
\ . \label{eqhm}
\end{equation}
The latter is equivalent to the claim (\ref{hsmub}). Let us
proceed to the R\'{e}nyi formulation. We first note that
\begin{equation}
R_{\alpha}(\clb_{m}|\bro)+R_{\beta}(\clb_{n}|\bro)\geq\ln{d}
\ . \label{rrab}
\end{equation}
It easily follows from the above relation between norm-like
functionals of the probability distributions
\cite{birula06,rast10r,rast104}. Applying the same reasons to
equation (\ref{rrab}), we have arrived at the claim (\ref{rsmub}).
$\blacksquare$

The inequalities (\ref{hsmub}) and (\ref{rsmub}) respectively
give lower bounds on symmetrized Tsallis and Renyi entropies
averaged over the set of several MUBs. In the particular case
$s=1$, both the inequalities are reduced to the bound on the
averaged Shannon entropy. This lower bound of Maassen--Uffink type
was previously derived in reference \cite{ballester}. Here, we
have obtained its extension to some of generalized entropies.

\section{Index of coincidence for a SIC-POVM}\label{sec04}

In this section, we calculate the index of coincidence for a
single SIC-POVM. The results are interested in own rights and also
for deriving state-dependent uncertainty relations for SIC-POVMs.
Let us fix some orthonormal basis in $\hh$. To each vector
$|\psi\rangle\in\hh$, we define $|\psi^{*}\rangle\in\hh$ as the
vector such that its components are conjugate to the corresponding
components of the $|\psi\rangle$. So, for any two vectors
$|\varphi\rangle$ and $|\psi\rangle$ we have
\begin{equation}
\langle\varphi^{*}|\psi^{*}\rangle=\langle\varphi|\psi\rangle^{*}=\langle\psi|\varphi\rangle
\ . \label{trvl}
\end{equation}
Our calculations are essentially based on the completeness
relation for a SIC-POVM, namely
\begin{equation}
\frac{1}{d}{\,}\sum_{j=1}^{d^{2}} |\phi_{j}\rangle\langle\phi_{j}|=\pen
\ . \label{compr1}
\end{equation}
We first establish an auxiliary result. The following statement
takes place.

\newtheorem{aropn1}[lm1]{Lemma}
\begin{aropn1}\label{pnn1}
Let $\omega$ be a primitive $d^{2}$th root of unity.
For the given $d^{2}$ unit vectors $|\phi_{j}\rangle$, we
introduce $d^{2}$ vectors of the space $\hh\otimes\hh$, namely
\begin{align}
|\Phi\rangle&=\frac{1}{d^{3/2}}{\,}\sum_{j=1}^{d^{2}} {\,}|\phi_{j}\rangle\otimes|\phi_{j}^{*}\rangle
\ , \label{phdf}\\
|\Psi_{k}\rangle&=\frac{\sqrt{d+1}}{d^{3/2}}{\,}\sum_{j=1}^{d^{2}} {\,}\omega^{k(j-1)}|\phi_{j}\rangle\otimes|\phi_{j}^{*}\rangle
\ , \label{pskdf}
\end{align}
where $k=1,\ldots,d^{2}-1$. If the unit
kets $|\phi_{j}\rangle$ satisfy equation (\ref{compr1}), then the
$d^{2}$ vectors (\ref{phdf})--(\ref{pskdf}) form an orthonormal
basis in the space $\hh\otimes\hh$.
\end{aropn1}

{\bf Proof.} First, we aim to show that the vectors
(\ref{phdf})--(\ref{pskdf}) are mutually orthogonal. The inner
product $\langle\Phi|\Psi_{k}\rangle$ is expressed, up to a
factor, as the sum
\begin{equation}
\sum_{i=1}^{d^{2}} \sum_{j=1}^{d^{2}} {\,}\omega^{k(j-1)} \bigl|\langle\phi_{i}|\phi_{j}\rangle\bigr|^{2}
=\sum_{j=1}^{d^{2}} \omega^{k(j-1)}+\sum_{\substack{i,j=1 \\ i\neq{j}}}^{d^{2}} \frac{\omega^{k(j-1)}}{d+1}
\ . \label{phps1}
\end{equation}
Here, we used formulas (\ref{undn1}) and (\ref{trvl}). As $\omega$
is a primitive root of unity, $\sum_{j=1}^{d^{2}}
\omega^{k(j-1)}=0$ for $k=1,\ldots,d^{2}-1$. In the right-hand
side of equation (\ref{phps1}), we multiply this zero sum by
factor $(d+1)^{-1}$ and obtain
\begin{equation}
\frac{1}{d+1}{\,}\sum_{i=1}^{d^{2}}\sum_{j=1}^{d^{2}} {\,}\omega^{k(j-1)}
=\frac{d^{2}}{d+1}{\,}\sum_{j=1}^{d^{2}} {\,}\omega^{k(j-1)}
\ . \label{phps2}
\end{equation}
The latter is zero for all $k=1,\ldots,d^{2}-1$. Further, we write
the inner product $\langle\Psi_{q}|\Psi_{k}\rangle$ in the form
\begin{align}
&\frac{d+1}{d^{3}}{\,}\sum_{i=1}^{d^{2}}
\sum_{j=1}^{d^{2}}{\,}\omega^{-q(i-1)} \omega^{k(j-1)}
\bigl|\langle\phi_{i}|\phi_{j}\rangle\bigr|^{2}=
\nonumber\\
&\frac{d+1}{d^{3}}{\,}\sum_{j=1}^{d^{2}}{\,}\omega^{(k-q)(j-1)}+
\frac{1}{d^{3}}\sum_{\substack{i,j=1 \\ i\neq{j}}}^{d^{2}}{\,}\omega^{-q(i-1)+k(j-1)}
{\>}. \label{psps1}
\end{align}
For $q\neq{k}$, the first sum in the right-hand side of equation
(\ref{psps1}) is zero. Multiplying this sum by $(d+1)^{-1}$, we
add it to the second sum in the right-hand side of equation
(\ref{psps1}) and get
\begin{equation}
\frac{1}{d^{3}}{\,}\sum_{i=1}^{d^{2}}{\,}\omega^{-q(i-1)} \sum_{j=1}^{d^{2}}{\,}\omega^{k(j-1)}=0
\ . \label{psps2}
\end{equation}
Let us proceed to the normalization. The squared norm of the
vector (\ref{phdf}) is equal to
\begin{equation}
\frac{1}{d^{3}}{\,}\sum_{i=1}^{d^{2}} \sum_{j=1}^{d^{2}}{\,}\bigl|\langle\phi_{i}|\phi_{j}\rangle\bigr|^{2}
=\frac{1}{d^{3}}\left(d^{2}+\frac{d^{4}-d^{2}}{d+1}\right)=1
\ . \label{phph1}
\end{equation}
Taking $q=k$ in equation (\ref{psps1}), the squared norm of the
vector (\ref{pskdf}) is written as
\begin{equation}
\frac{d+1}{d^{3}}\biggl\{d^{2}\!\left(1-\frac{1}{d+1}\right)
+\frac{1}{d+1}{\,}\sum_{i=1}^{d^{2}}{\,}\omega^{-k(i-1)}\sum_{j=1}^{d^{2}}{\,}\omega^{k(j-1)}\biggr\}=1
\ , \label{psps12}
\end{equation}
since the second sum in the left-hand side vanishes.
$\blacksquare$

So, each SIC-POVM in $d$-dimensional space $\hh$ can be converted
into the orthonormal basis in the space $\hh\otimes\hh$. In some
respects, our construction (\ref{phdf})--(\ref{pskdf}) is similar
to the vectors built of several MUBs in reference \cite{molm09}.
Using such vectors, Wu et al. \cite{molm09} derived
state-dependent uncertainty relations for any set of MUBs in terms
of the Shannon and Renyi $2$-entropies. We have already extended
these relations to the Tsallis $\alpha$-entropies for
$\alpha\in(0;2]$ and the R\'{e}nyi $\alpha$-entropies for
$\alpha\in[2;\infty)$. In reference \cite{molm09}, the basic point
is an upper bound on the corresponding sum of the indices of
coincidence. In the following, we will calculate exactly the index
of coincidence for any SIC-POVM and arbitrary density matrix.

\newtheorem{propn2}[prop1]{Proposition}
\begin{propn2}\label{pnn2}
Let $\nc=\left\{d^{-1}|\phi_{j}\rangle\langle\phi_{j}|\right\}$ be
a SIC-POVM in $d$-dimensional Hilbert space $\hh$. For arbitrary
density matrix $\bro$ on $\hh$, the generated probability
distribution satisfies
\begin{equation}
\sum_{j=1}^{d^{2}} p_{j}(\nc|\bro)^{2}=\frac{\tr(\bro^{2})+1}{d(d+1)}
\ . \label{indc0}
\end{equation}
\end{propn2}

{\bf Proof.} We first note two immediate consequences of the
completeness relation (\ref{compr1}) for a SIC-POVM. For arbitrary
operator $\am\in\lnp(\hh)$ and vector $|\psi\rangle\in\hh$, one
gives
\begin{align}
& \frac{1}{d^{2}}{\,}\sum_{i=1}^{d^{2}}\sum_{j=1}^{d^{2}}{\,}\langle\phi_{i}|\am|\phi_{j}\rangle\langle\phi_{j}|\phi_{i}\rangle=\tr(\am)
\ , \label{con1}\\
& \frac{1}{d}{\,}\sum_{j=1}^{d^{2}}{\,}|\phi_{j}\rangle\langle\phi_{j}|\psi\rangle=|\psi\rangle
\ , \label{con2}
\end{align}
These results are yielded by substituting formula (\ref{compr1})
into the identities $\tr(\pen\am\pen)=\tr(\am)$ and
$\pen{\,}|\psi\rangle=|\psi\rangle$, respectively. Let us consider
the vector $\bro\otimes\pen|\Phi\rangle\in\hh\otimes\hh$. It can
be represented as a linear combination of the basis vectors
$|\Phi\rangle$ and $|\Psi_{k}\rangle$ defined in equations
(\ref{phdf})--(\ref{pskdf}). Using equation (\ref{con1}) and the
normalization $\tr(\bro)=1$, we have
\begin{equation}
\langle\Phi|\bro\otimes\pen|\Phi\rangle
=\frac{1}{d^{3}}{\,}\sum_{i=1}^{d^{2}}\sum_{j=1}^{d^{2}}{\,}\langle\phi_{i}|\bro|\phi_{j}\rangle\langle\phi_{j}|\phi_{i}\rangle
=\frac{1}{d}
\ . \label{bpph}
\end{equation}
Hence, the vector $\bro\otimes\pen|\Phi\rangle$ is represented in
the form
\begin{equation}
\bro\otimes\pen|\Phi\rangle=\frac{1}{d}{\,}|\Phi\rangle+\sum_{k=1}^{d^{2}-1}{\,}a_{k}|\Psi_{k}\rangle
\ . \label{exph}
\end{equation}
The coefficients
$a_{q}=\langle\Psi_{q}|\bro\otimes\pen|\Phi\rangle$ are calculated
as
\begin{equation}
a_{q}=\frac{\sqrt{d+1}}{d^{3}}{\,}\sum_{i=1}^{d^{2}}\sum_{j=1}^{d^{2}}{\,}
\omega^{-q(i-1)}\langle\phi_{i}|\bro|\phi_{j}\rangle\langle\phi_{j}|\phi_{i}\rangle
=\frac{\sqrt{d+1}}{d^{2}}{\,}\sum_{i=1}^{d^{2}}{\,}\omega^{-q(i-1)}\langle\phi_{i}|\bro|\phi_{i}\rangle
=\frac{\sqrt{d+1}}{d}{\,}\sum_{i=1}^{d^{2}}{\,}\omega^{-q(i-1)}p_{i}
\ , \label{akps}
\end{equation}
where we used equation (\ref{con2}) and
$p_{i}=d^{-1}\langle\phi_{i}|\bro|\phi_{i}\rangle$. The squared
norm of the vector $\bro\otimes\pen|\Phi\rangle$ can be expressed
in two different ways. Due to formula (\ref{con1}), we write the
first expression
\begin{equation}
\langle\Phi|(\bro\otimes\pen)^{2}|\Phi\rangle
=\frac{1}{d^{3}}{\,}\sum_{i=1}^{d^{2}}\sum_{j=1}^{d^{2}}{\,}\langle\phi_{i}|\bro^{2}|\phi_{j}\rangle\langle\phi_{j}|\phi_{i}\rangle
=\frac{1}{d}{\>}\tr(\bro^{2})
\ . \label{fsex}
\end{equation}
The second expression is obtained from equation (\ref{exph}) as
the sum of squared modula of the coefficients. Namely, we obtain
\begin{equation}
\frac{1}{d^{2}}+\sum_{k=1}^{d^{2}-1}{\,}a_{k}^{*}a_{k}
=\frac{1}{d^{2}}
+\frac{d+1}{d^{2}}{\,}\sum_{i=1}^{d^{2}}{\,}\sum_{j=1}^{d^{2}}{\,}p_{i}p_{j}\sum_{k=1}^{d^{2}-1}{\,}\omega^{k(i-j)}
=\frac{1}{d^{2}}+(d+1)\sum_{j=1}^{d^{2}}{\,}p_{j}^{2}-\frac{d+1}{d^{2}}
\ . \label{snex}
\end{equation}
At the last step, we use the formulas
$\sum_{k=1}^{d^{2}-1}\omega^{k(i-j)}=d^{2}\delta_{ij}-1$ and
$\sum_{j=1}^{d^{2}}p_{j}=1$. Combining equations (\ref{fsex}) and
(\ref{snex}) leads to the formula
\begin{equation}
\frac{1}{d}{\,}\tr(\bro^{2})=(d+1)\sum_{j=1}^{d^{2}}{\,}p_{j}^{2}-\frac{1}{d}
\ , \label{fssn}
\end{equation}
which is equivalent to the claim (\ref{indc0}).
$\blacksquare$

Thus, for any SIC-POVM the index of coincidence of generated
probability distribution is exactly expressed in terms of measured
density matrix. In general, probabilities will range in the domain
defined by the conditions $p_{j}\geq0$ for all $j$ and
$\sum_{j}p_{j}=1$. In the case of SIC-POVMs, they should also
satisfy the additional restriction (\ref{indc0}), which depends on
measured state. On the other hand, measurement statistics can be
used in studying quantum states. Using full statistics of a
SIC-POVM and the result (\ref{indc0}), we can exactly find the
trace of square of unknown density matrix. This trace is one of
common measures of a degree of state impurity. For a pure state
$\bro=|\psi\rangle\langle\psi|$, the result (\ref{indc0}) is
reduced to
\begin{equation}
\sum_{j=1}^{d^{2}} p_{j}(\nc|\psi)^{2}=\frac{2}{d(d+1)}
\ . \label{indc0p}
\end{equation}
That is, the index of coincidence is constant for all pure states.
This claim has been proved in reference \cite{rottler} as a
corollary of the fact that the kets $|\phi_{j}\rangle$ form a
spherical $2$-design. When $|\psi\rangle$ is taken as one of the
kets $|\phi_{j}\rangle$, the result (\ref{indc0p}) directly
follows from equation (\ref{undn1}). Another easy example is
provided with the completely mixed state $\bro_{*}=\pen/d$. For a
SIC-POVM, we have the equiprobable distribution, i.e.
$p_{j}=1/d^{2}$ for all $j=1,\ldots,d^{2}$. In this example, the
index of coincidence is equal to
\begin{equation}
\sum_{j=1}^{d^{2}} p_{j}(\nc|\bro_{*})^{2}=\frac{1}{d^{2}}
\ . \label{indcm}
\end{equation}
Substituting $\tr\bigl(\bro_{*}^{2}\bigr)=1/d$, the right-hand
side of equation (\ref{indc0}) is reduced to the right-hand side
of equation (\ref{indcm}). Thus, formula (\ref{indc0}) is an
extension of the important result (\ref{indc0p}) to all density
matrices. We will use equations (\ref{indc0}) and (\ref{indc0p})
in deriving uncertainty relations for a SIC-POVM in terms of both
the R\'{e}nyi and Tsallis entropies. In some respects, our
approach is similar to the proof of theorem 1 of reference
\cite{molm09}. For a set of MUBs, this theorem gives an upper
bound on the corresponding sum of the indices of coincidence. For
a SIC-POVM, however, the index of coincidence enjoys just the
equality, instead of some kind of inequality. This very
interesting result is a manifestation of the symmetric structure
of SIC-sets. Together with the normalization condition, the
generated probabilities $p_{j}(\nc|\bro)$ should also obey formula
(\ref{indc0}) for all density matrices. Such a conclusion concurs
with the fact that SIC-POVMs are difficult to construct. In this
regard, the authors of reference \cite{krsw05} considered
approximate versions, which allow small deviation from the
condition (\ref{undn1}). It would be interesting to study further
changes induced by such a deviation.

In the case $d=2$, the relation (\ref{indc0}) can be represented
in terms of the Bloch vector. This description is commonly used in
state reconstruction \cite{gparis06}. In this case, we will denote
the identity $2\times2$-matrix by $\pen$ and the usual Pauli
matrices by $\bsg_{x}$, $\bsg_{y}$, and $\bsg_{z}$, respectively.
It is convenient to write a qubit density matrix in the form
\begin{equation}
\bro=\frac{1}{2}{\>}\bigl(\pen+\Vec{s}\cdot\Vec{\bsg}\bigr)
{\>}. \label{bvr2}
\end{equation}
where $\Vec{s}=(s_{x},s_{y},s_{z})$ is the Bloch vector of $\bro$.
The matrix is obviously of unit trace. Positivity of this matrix
is provided by the condition $s=|\Vec{s}|\leq1$. Since
$\tr(\bro^{2})=(1+s^{2})/2$, the index of coincidence is
written as
\begin{equation}
\sum_{j=1}^{4} p_{j}(\nc|\bro)^{2}=\frac{3+s^{2}}{12}
\ . \label{indcd2}
\end{equation}
For all pure states, we have $s=1$ and the number $1/3$ in
equation (\ref{indcd2}). For the completely mixed state, the index
of coincidence is equal to $1/4$ due to $s=0$. The more a state is
mixed, the less the index (\ref{indcd2}). A similar conclusion
holds in higher dimensions. Moreover, for large $d$ the pure-state
value (\ref{indc0p}) is almost the doubled value (\ref{indcm}).

The new result (\ref{indc0}) is somewhat similar to the result
(\ref{lwbm}) for MUBs given in reference \cite{molm09}. In
reference \cite{shbah12}, the bound (\ref{lwbm}) has been used in
studying entanglement detection via MUBs. In principle, the exact
relation (\ref{indc0}) may also be applied in this context. We
consider a bipartite system of two $d$-dimensional subsystems. Its
Hilbert space is the product $\hh_{AB}=\hh_{A}\otimes\hh_{B}$ of
two isomorphic spaces $\hh_{A}$ and $\hh_{B}$. Fixing some
orthonormal basis $\bigl\{|n\rangle\bigr\}$, a maximally entangled
pure state is written as
\begin{equation}
|\Phi_{+}\rangle=\frac{1}{\sqrt{d}}\sum_{n=1}^{d} |n\rangle\otimes|n\rangle
\ . \label{mest}
\end{equation}
To a SIC-POVM
$\nc_{A}=\bigl\{d^{-1}|\phi_{i}\rangle\langle\phi_{i}|\bigr\}$ on
$\hh_{A}$, we assign
$\nc_{B}=\bigl\{d^{-1}|\phi_{j}^{*}\rangle\langle\phi_{j}^{*}|\bigr\}$
on $\hh_{B}$. They give the POVM $\nc_{AB}$ on $\hh_{AB}$ with
elements
\begin{equation}
\widetilde{\nm}_{ij}=\frac{1}{d^{2}}{\>}|\phi_{i}\phi_{j}^{*}\rangle\langle\phi_{i}\phi_{j}^{*}|
\ , \qquad
|\phi_{i}\phi_{j}^{*}\rangle\equiv|\phi_{i}\rangle\otimes|\phi_{j}^{*}\rangle
\ . \label{abpm}
\end{equation}
Generally, it is not a SIC-POVM on $\hh_{AB}$. Let $\wbro$ be a
density matrix of the bipartite system. Using probabilities
$P(i,j)={\tr}{\bigl(\widetilde{\nm}_{ij}\wbro\bigr)}$, we can put
a correlation measure
\begin{equation}
G(\nc_{AB}|\wbro)=\sum_{j=1}^{d^{2}}P(j,j)
\ . \label{cmjj}
\end{equation}
This measure is similar to the mutual predictability proposed in
reference \cite{shbah12}. For each state of the form
$\wbro=\bro_{A}\otimes\bro_{B}$, the probability $P(i,j)$ is
merely a product of two particular probabilities. Using equation
(\ref{indc0}) and the Cauchy--Schwarz inequality, we then obtain
\begin{align}
G(\nc_{AB}|\bro_{A}\otimes\bro_{B})
&\leq\frac{\sqrt{{\tr}{\bigl(\bro_{A}^{2}\bigr)}+1}\sqrt{{\tr}{\bigl(\bro_{B}^{2}\bigr)}+1}}{d(d+1)}
\label{gab1}\\
&\leq\frac{2}{d(d+1)}
\ . \label{gab2}
\end{align}
All separable states obey the condition (\ref{gab2}). Let us compare this bound with the value of correlation
measure for the entangled state (\ref{mest}). As each
ket $|\phi_{j}\rangle$ is unit, we have
\begin{equation}
\langle\phi_{j}\phi_{j}^{*}|\Phi_{+}\rangle
=\frac{1}{\sqrt{d}}\sum_{n=1}^{d} \bigl|\langle\phi_{j}|n\rangle\bigr|^{2}
=\frac{1}{\sqrt{d}}
\end{equation}
Hence, we have the diagonal probabilities $P(j,j)=d^{-3}$ and the
value $G(\nc_{AB}|\Phi_{+})=d^{-1}$. In  high dimensions, the
latter is significantly larger than the upper bound (\ref{gab2}).
Of course, the bound (\ref{gab2}) for separable states gives only
a necessary criterion. Here, we do not aim to address a very
complicated problem of entanglement detection in details. We
rather wish to motivate that the results (\ref{indc0}) and
(\ref{indc0p}) may be useful in such studies.

\section{Lower entropic bounds for SIC-POVMs}\label{sec4}

In this section, we derive entropic inequalities for symmetric
informationally complete POVMs. First, we obtain lower bounds for
a single SIC-POVM in terms of both the Tsallis  and R\'{e}nyi
entropies. Second, entropic bounds of Maassen--Uffink type for a
pair of SIC-POVMs are given.

\subsection{Lower entropic bounds for a single SIC-POVM}\label{sc41}

A simple way to obtain lower bounds on the Shannon entropy of a
given POVM was noticed by Massar \cite{massar07}. Developing this
idea with the right-hand side of equation (\ref{tsaent2}), for all
$\alpha>0$ we write
\begin{equation}
H_{\alpha}(\nc|\bro)\geq
{\ln_{\alpha}}{\left(\frac{1}{\max{p}_{j}}\right)}\geq\ln_{\alpha}(d)
\ . \label{masd}
\end{equation}
Indeed, for a SIC-POVM we have
$p_{j}=d^{-1}\langle\phi_{j}|\bro|\phi_{j}\rangle\leq{d}^{-1}$ for
all $j$. Further, the function $x\mapsto\ln_{\alpha}(1/x)$
decreases with $x$ for all $\alpha>0$. In the R\'{e}nyi case, we
have the relation
\begin{equation}
R_{\alpha}(\nc|\bro)\geq-\ln(\max{p}_{j})\geq\ln{d}
\ . \label{mard}
\end{equation}
To prove this, we note that the function
$x\mapsto(1-\alpha)^{-1}\ln{x}$ increases for $0<\alpha<1$ and
decreases for $1<\alpha<\infty$. Combining this with the relation
\begin{equation}
\sum\nolimits_{j} p^{\alpha}
\left\{
\begin{array}{cc}
\geq, & 0<\alpha<1 \\
\leq, & 1<\alpha<\infty
\end{array}
\right\}
(\max{p}_{j})^{\alpha-1}\sum\nolimits_{j} p_{j}
\ , \label{aa01}
\end{equation}
we then get the claim (\ref{mard}). The right-hand side of
equation (\ref{mard}) is actually an immediate lower bound on the
min-entropy, when $\alpha=\infty$. Its validity for other orders
also follows from the fact that R\'{e}nyi's $\alpha$-entropy is a
non-increasing function of $\alpha$. In both the formulations,
more stronger entropic bound can be obtained from equations (\ref{indc0}) and
(\ref{indc0p}). The following statement takes place.

\newtheorem{prop4}[prop1]{Proposition}
\begin{prop4}\label{pan4}
Let $\nc$ be a SIC-POVM in $d$-dimensional Hilbert space $\hh$.
For $\alpha\in(0;2]$ and each density matrix $\bro$ on $\hh$,
the Tsallis $\alpha$-entropy satisfies the state-dependent bound
\begin{equation}
H_{\alpha}(\nc|\bro)\geq
{\ln_{\alpha}}{\left(\frac{d(d+1)}{\tr(\bro^{2})+1}\right)}
{\,}. \label{aslb1}
\end{equation}
\end{prop4}

{\bf Proof.} For $\alpha\in(0;2]$ and any probability
distribution, we have proved equation (\ref{conc1}). Combining
this with equation (\ref{indc0}) immediately gives the result
(\ref{aslb1}).
$\blacksquare$

For pure states, we have $\tr(\bro^{2})=1$. In this case, the
right-hand side of equation (\ref{aslb1}) is reduced to the
state-independent form
\begin{equation}
H_{\alpha}(\nc|\bro)\geq
{\ln_{\alpha}}{\left(\frac{d(d+1)}{2}\right)}
{\,}, \label{aslb0}
\end{equation}
which actually holds for any state. In the standard case
$\alpha=1$, the uncertainty relation (\ref{aslb1}) gives a lower
bound on the Shannon entropy, namely
\begin{equation}
H_{1}(\nc|\bro)\geq{\ln}{\left(\frac{d(d+1)}{\tr(\bro^{2})+1}\right)}
\geq{\ln}{\left(\frac{d(d+1)}{2}\right)}
{\,}. \label{scaslb}
\end{equation}
It is stronger than the inequality
$H_{1}(\nc|\bro)\geq\ln{d}$, which follows from formula
(\ref{masd}). Moreover, for large $d$ the bound (\ref{scaslb}) is
almost $2\ln{d}$. For enough high dimensions and order $\alpha$
around $1$, therefore, the uncertainty relation (\ref{aslb1}) is
significantly stronger than the simple bound (\ref{masd}). On the
other hand, the lower bound (\ref{masd}) holds for all $\alpha>0$.

Let us consider a case of detection inefficiencies. We will again
use the model connected with equations (\ref{dspd}) and
(\ref{qtlm0}). Similarly to equation (\ref{dspd1}), the actual
probabilities are assumed to be distorted as
\begin{equation}
p_{j}^{(\eta)}(\nc|\bro)=\eta{\,}p_{j}(\nc|\bro)
\ , \qquad
p_{\varnothing}^{(\eta)}(\nc|\bro)=1-\eta
\ . \label{nnpd1}
\end{equation}
Here, the parameter $\eta\in[0;1]$ describes a detector
efficiency. In equation (\ref{nnpd1}), the second quantity is the
no-click probability. Due to equations (\ref{qtlm0}) and
(\ref{aslb1}), the entropy $H_{\alpha}^{(\eta)}(\nc|\bro)$ of the
distribution (\ref{nnpd1}) obeys
\begin{equation}
H_{\alpha}^{(\eta)}(\nc|\bro)\geq
\eta^{\alpha}{\,}{\ln_{\alpha}}{\left(\frac{d(d+1)}{\tr(\bro^{2})+1}\right)}
+h_{\alpha}(\eta)
{\ }, \label{aslb1et}
\end{equation}
where $\alpha\in(0;2]$. The result (\ref{aslb1et}) is an entropic
uncertainty relation for a single SIC-POVM in the model of
detection inefficiencies. In the standard case $\alpha=1$, the
inefficiency-free lower bound (\ref{scaslb}) is merely added by
the binary Shannon entropy $h_{1}(\eta)$. Thus, an additional
uncertainty is introduced by the detector. To estimate a required
amount of detector efficiency, we demand the following. The first
term of the right-hand side of equation (\ref{aslb1et}) should be
sufficiently large in comparison with the second. We now consider
the R\'{e}nyi formulation. It directly follows from equations
(\ref{rain}) and (\ref{indc0}).

\newtheorem{prop5}[prop1]{Proposition}
\begin{prop5}\label{pan5}
Let $\nc$ be a SIC-POVM in $d$-dimensional Hilbert space $\hh$.
For $\alpha\in[2;\infty)$ and each density matrix $\bro$ on
$\hh$, the R\'{e}nyi $\alpha$-entropy satisfies the
state-independent bound
\begin{equation}
R_{\alpha}(\nc|\bro)\geq
\frac{\alpha}{2(\alpha-1)}{\>}{\ln}{\left(\frac{d(d+1)}{\tr(\bro^{2})+1}\right)}
{\,}. \label{aslr1}
\end{equation}
\end{prop5}

Since $\tr(\bro^{2})\leq1$, we get the
state-independent bound
\begin{equation}
R_{\alpha}(\nc|\bro)\geq
\frac{\alpha}{2(\alpha-1)}{\>}{\ln}{\left(\frac{d(d+1)}{2}\right)}
{\,}. \label{aslr0}
\end{equation}
In the range $\alpha\in[2;\infty)$, we herewith have the lower
bound (\ref{aslr1}), which depends on $\alpha$. For $\alpha=2$, the
inequality (\ref{aslr1}) gives the lower bound on the collision
entropy, namely
\begin{equation}
R_{2}(\nc|\bro)\geq{\ln}{\left(\frac{d(d+1)}{\tr(\bro^{2})+1}\right)}
{\,}. \label{aslr2}
\end{equation}
Since the R\'{e}nyi $\alpha$-entropy does not increase with
$\alpha$, this bound holds for all Renyi's entropies of order
$\alpha\in(0;2]$. In this range, we have only the constant
entropic bound. Further, the right-hand side of equation
(\ref{aslr1}) clearly decreases with $\alpha$, since the logarithm
multiplier is $(1/2)\bigl(1+(\alpha-1)^{-1}\bigr)$. So, for large
values of $\alpha$ the lower bound (\ref{aslr1}) is approximately
a half of its size for $\alpha=2$. In the limit $\alpha\to\infty$,
formula (\ref{aslr1}) leads to the min-entropy relation
\begin{equation}
R_{\infty}(\nc|\bro)\geq
\frac{1}{2}{\>}{\ln}{\left(\frac{d(d+1)}{\tr(\bro^{2})+1}\right)}
\geq\frac{1}{2}{\>}{\ln}{\left(\frac{d(d+1)}{2}\right)}
{\,}. \label{aslrin}
\end{equation}
Since $d(d+1)<2d^{2}$ for $d\geq2$, the right-hand side of equation
(\ref{aslrin}) is weaker than the bound (\ref{mard}). It is a
manifestation of the fact that the uncertainty relation
(\ref{aslr1}) is not tight. This observation gives a reason for
improving the min-entropy relation (\ref{aslrin}). Using the
result (\ref{aust}), we are in position to derive an improved
state-dependent relation for the min-entropy of a SIC-POVM.

\newtheorem{propn6}[prop1]{Proposition}
\begin{propn6}\label{pnn6}
Let $\nc$ be a SIC-POVM in $d$-dimensional Hilbert space $\hh$.
For each density matrix $\bro$ on $\hh$, the min-entropy of
generated probability distribution satisfies
\begin{equation}
R_{\infty}(\nc|\bro)\geq2\ln{d}-{\ln}{\left(1+\sqrt{d-1}\sqrt{\tr(\bro^{2}){\,}d-1}\right)}
{\,}. \label{reust}
\end{equation}
\end{propn6}

{\bf Proof.} Replacing $b^{2}$ with the term (\ref{indc0}) and $n$ with
$d^{2}$, the inequality (\ref{aust}) finally gives
\begin{equation}
\max\Bigl\{p_{j}(\nc|\bro):{\>}1\leq{j}\leq{d}^{2}\Bigr\}\leq
\frac{1}{d^{2}}\left(1+\sqrt{d-1}\sqrt{\tr(\bro^{2}){\,}d-1}\right)
{\,}. \label{reust1}
\end{equation}
The claim (\ref{reust}) is easily obtained from equations
(\ref{mnen}) and (\ref{reust1}).
$\blacksquare$

Proposition \ref{pnn6} provides an uncertainty relation for a
SIC-POVM in terms of the min-entropy. For a pure state
$\bro=|\psi\rangle\langle\psi|$, the right-hand side of
(\ref{reust}) is reduced to the right-hand side of (\ref{mard}).
The state-dependent relation (\ref{reust}) is obviously saturated,
whenever the state $|\psi\rangle$ is one of the kets
$|\phi_{j}\rangle$ comprising the SIC-POVM. Indeed, one of the
probabilities is then $1/d$ and other are all
$\bigl(d(d+1)\bigr)^{-1}$. Further, the inequality (\ref{reust})
is saturated with the completely mixed state $\bro_{*}=\pen/d$,
when $p_{j}=1/d^{2}$ for all $j$. The right-hand side of equation
(\ref{reust}) increases as $\tr(\bro^{2})$ decreases. In other
words, the more a state is mixed, the more the bound
(\ref{reust}). Thus, the uncertainty relation (\ref{reust}) enjoys
a natural dependence on states. It is also tight in several
certain cases. New entropic bound is essentially based on the
exact result (\ref{indc0}). As SIC-POVMs are hard to construct,
their approximate versions with a deviation from equation
(\ref{undn1}) are of interest \cite{krsw05}. Extensions of the
above relations to approximate SIC-POVMs would be a subject of
separate research. In principle, Lemma \ref{lapp2} could be used
for deriving a lower bound on the min-entropy in other cases. For
this purpose, the index of coincidence should be evaluated from
above.

\subsection{Entropic relations of Maassen--Uffink type}\label{sc42}

In general, entropic uncertainty relations for a pair of SIC-POVMs
immediately follow from results given in the literature. For two
POVMs, a state-independent bound on the sum of Shannon entropies
was derived in reference \cite{krpr02}. This bound was first
proved for projective measurements. Using Naimark's extension
step-by-step, the writers of reference \cite{krpr02} then
generalized this result to arbitrary POVMs. Reformulations with
state-dependent bounds or generalized entropic functions are of
interest. In references
\cite{rast10r,rast104,rast12num,rast12quasi}, we have developed
this issue in both directions. In the present paper, uncertainty
relations for rank-one POVMs will be sufficient. We include
another proof for convenience, in the hope that it may offer
additional insights.

\newtheorem{prop6}[prop1]{Proposition}
\begin{prop6}\label{pan6}
Let $\mc=\bigl\{|m_{i}\rangle\langle{m}_{i}|\bigr\}$ and
$\nc=\bigl\{|n_{j}\rangle\langle{n}_{j}|\bigr\}$ be two POVMs with
rank-one elements in $d$-dimensional Hilbert space $\hh$. To any
density matrix $\bro$, we assign the quantity
\begin{equation}
g\bigl(\mc,\nc|\bro\bigr):=\max\biggl\{
\frac{\bigl|\langle{m}_{i}|n_{j}\rangle{\,}\langle{n}_{j}|\bro|m_{i}\rangle\bigr|}
{\langle{m}_{i}|\bro|m_{i}\rangle^{1/2}{\,}\langle{n}_{j}|\bro|n_{j}\rangle^{1/2}}:
\langle{m}_{i}|\bro|m_{i}\rangle\neq0,{\>}\langle{n}_{j}|\bro|n_{j}\rangle\neq0
\biggr\}
{\,}. \label{gmndf}
\end{equation}
For $1/\alpha+1/\beta=2$ and $\mu=\max\{\alpha,\beta\}$, the
Tsallis entropies and R\'{e}nyi entropies respectively satisfy
\begin{align}
H_{\alpha}(\mc|\bro)+H_{\beta}(\nc|\bro)&\geq{\ln_{\mu}}{\left(g\bigl(\mc,\nc|\bro\bigr)^{-2}\right)}
{\,}, \label{onrun}\\
R_{\alpha}(\mc|\bro)+R_{\beta}(\nc|\bro)&\geq-2\ln{g}\bigl(\mc,\nc|\bro\bigr)
{\,}. \label{ornun}
\end{align}
\end{prop6}

{\bf Proof.} Using the completeness relation (\ref{cmprl}), the
$i$th probability of the measurement $\mc$ is rewritten as
\begin{equation}
q_{i}(\mc|\bro)=\langle{m}_{i}|\pen{\,}\bro|m_{i}\rangle=
\sum\nolimits_{j}\langle{m}_{i}|n_{j}\rangle{\,}\langle{n}_{j}|\bro|m_{i}\rangle
\ . \label{qipj0}
\end{equation}
Let us put the numbers
$v_{i}^{\prime}=\langle{m}_{i}|\bro|m_{i}\rangle^{1/2}$ and
$u_{j}^{\prime}=p_{j}(\nc|\bro)^{1/2}=\langle{n}_{i}|\bro|n_{j}\rangle^{1/2}$.
Dividing equation (\ref{qipj0}) by $v_{i}^{\prime}$, we then
obtain
\begin{align}
v_{i}^{\prime}&=\sum\nolimits_{j}t_{ij}{\,}u_{j}^{\prime}
\ , \label{qipj01}\\
t_{ij}&=
\frac{\langle{m}_{i}|n_{j}\rangle{\,}\langle{n}_{j}|\bro|m_{i}\rangle}{\langle{m}_{i}|\bro|m_{i}\rangle^{1/2}{\,}\langle{n}_{j}|\bro|n_{j}\rangle^{1/2}}
\ . \label{qipj1}
\end{align}
That is, the output tuple $v^{\prime}$ is obtained from the input
$u^{\prime}$ by a linear transformation. To apply the Riesz
theorem, we should beforehand check the precondition (\ref{prcp}).
It is required that the latter be valid for arbitrary input tuple
$u$. The proof of this point is direct but somewhat technical (for
details, see Appendix \ref{appr}). From equation (\ref{spps}), we
further obtain the relation \begin{equation}
\|q\|_{\alpha}\leq{g}\bigl(\mc,\nc|\bro\bigr)^{2(1-\beta)/\beta}{\,}\|p\|_{\beta}
\ , \label{rnlf}
\end{equation}
in which $1/\alpha+1/\beta=2$ and $1/2<\beta<1$. As was shown in
section 3 of Ref. \cite{rast104}, this relation leads to the lower
entropic bounds (\ref{onrun}) and (\ref{ornun}). The former is
obtained by minimization of a certain function under equation
(\ref{rnlf}); the latter is directly obtained by taking the
logarithm of equation (\ref{rnlf}).
$\blacksquare$

The presented reasonings differ from the previous formulations as
follows. It is emphasized in the above proof that an extension to
orthogonal sets is actually required only in checking the
precondition (\ref{prcp}). The expression (\ref{qipj1}) for
transformation elements is immediately obtained in easy way. For
two SIC-POVMs
$\mc=\bigl\{d^{-1}|\varphi_{i}\rangle\langle\varphi_{i}|\bigr\}$
and $\nc=\bigl\{d^{-1}|\phi_{j}\rangle\langle\phi_{j}|\bigr\}$,
the function (\ref{gmndf}) reads
\begin{equation}
g\bigl(\mc,\nc|\bro\bigr)=\frac{1}{d}{\,}\max\biggl\{
\frac{\bigl|\langle\varphi_{i}|\phi_{j}\rangle{\,}\langle\phi_{j}|\bro|\varphi_{i}\rangle\bigr|}
{\langle\varphi_{i}|\bro|\varphi_{i}\rangle^{1/2}{\,}\langle\phi_{j}|\bro|\phi_{j}\rangle^{1/2}}:
\langle\varphi_{i}|\bro|\varphi_{i}\rangle\neq0,{\>}\langle\phi_{j}|\bro|\phi_{j}\rangle\neq0
\biggr\}
{\,}. \label{smndf}
\end{equation}
Substituting this expression into equations (\ref{onrun}) and
(\ref{ornun}) directly gives uncertainty relations for the
SIC-POVMs $\mc$ and $\nc$ in terms of Tsallis' and R\'{e}nyi's
entropies. It follows from the Cauchy--Schwarz inequality that
\begin{equation}
\bigl|\langle\phi_{j}|\bro|\varphi_{i}\rangle\bigr|\leq
\langle\varphi_{i}|\bro|\varphi_{i}\rangle^{1/2}{\,}\langle\phi_{j}|\bro|\phi_{j}\rangle^{1/2}
\ . \label{csin}
\end{equation}
Hence, the quantity (\ref{smndf}) obeys
\begin{equation}
g\bigl(\mc,\nc|\bro\bigr)
\leq\frac{1}{d}{\>}\underset{ij}{\max}\Bigl\{\bigl|\langle\varphi_{i}|\phi_{j}\rangle\bigr|\Bigr\}
=:\bar{f}(\mc,\nc)
\ . \label{fbar}
\end{equation}
Further, the functions $x\mapsto\ln_{\mu}\bigl(x^{-2}\bigr)$ and
$x\mapsto-2\ln{x}$ both decrease. From equations (\ref{onrun}) and
(\ref{ornun}), we then obtain the state-independent uncertainty
relations for two SIC-POVMs. For $1/\alpha+1/\beta=2$ and
$\mu=\max\{\alpha,\beta\}$, there holds
\begin{align}
H_{\alpha}(\mc|\bro)+H_{\beta}(\nc|\bro)&\geq{\ln_{\mu}}{\Bigl(\bar{f}(\mc,\nc)^{-2}\Bigr)}
{\,}, \label{fnrun}\\
R_{\alpha}(\mc|\bro)+R_{\beta}(\nc|\bro)&\geq-2\ln\bar{f}(\mc,\nc)
{\ }. \label{frnun}
\end{align}
These lower bounds are uncertainty relations of Maassen--Uffink
type. The right-hand sides of equation (\ref{fnrun}) and
(\ref{frnun}) are also lower bounds on the sum of corresponding
symmetrized entropies. Here, the entropic parameters are put in
accordance with equation (\ref{prms}).

\section{Conclusions}\label{sec5}

We have established entropic uncertainty relations for mutually
unbiased bases as well for symmetric informationally complete
POVMs. The R\'{e}nyi and Tsallis formulations are both presented.
The min-entropy uncertainty relations of state-dependent form are
separately considered. For a set of MUBs, we give the
state-dependent entropic bounds and their state-independent forms.
In terms of Tsallis $\alpha$-entropies, entropic uncertainty
relations are derived for all $\alpha\in(0;2]$. Bounds on the
corresponding R\'{e}nyi $\alpha$-entropies are obtained for all
$\alpha\in[2;\infty)$. We also derived bounds of
Maassen--Uffink type in term of the symmetrized
entropies. Our uncertainty relations for MUBs are an
extension of previous results to generalized entropies.

Further, we have obtained novel entropic bounds for SIC-POVMs.
They are based on the exactly calculated index of coincidence of
generated probability distribution. This calculation is a new
result of own significance. We briefly discussed it in the context
of entanglement detection. Entropic lower bounds of
state-independent form for a single SIC-POVM are obtained in terms
of both the R\'{e}nyi and Tsallis entropies in wide ranges of
parameters. Using the Tsallis entropies, we also considered
uncertainty relations with detection inefficiences. Entropic
uncertainty relations of Maassen--Uffink type for a pair of
SIC-POVMs were formulated as well. In this regard, we proposed a
short derivation of the uncertainty relation for two rank-one
POVMs.

Entropic uncertainty relations are not only another tool
for expressing some trade-off. Indeed, entropic bounds give
certain conditions on probabilities of corresponding measurements.
Hence, such relations may be of practical interest. State
preparation and measurements are essential steps in quantum
protocols. Keeping statistics of events, legitimate users will be
able to check proper restrictions on probabilities. When quantum
carriers are the subject of an external activity, actual
correlations will be altered somehow. In principle, this may lead
to a violation of  existing conditions. For such purposes,
legitimate users would like to have as many testable conditions as
possible. One of reasonable ways is provided by generalized
entropies. In general, this issue deserves further investigations.

\medskip
{\small{The author is grateful to Li Yu for useful
comments on the first version of this paper. The author also
thanks Christopher Fuchs and Karol \.{Z}yczkowski for helpful
correspondence.}}

\appendix

\section{An inequality}\label{astt}

In this appendix, we derive an inequality used in obtaining lower
bounds in terms of the min-entropy. The following statement takes
place.
\newtheorem{app2}[lm1]{Lemma}
\begin{app2}\label{lapp2}
Let $n$ positive numbers $x_{j}$ satisfy the two conditions
$\sum_{j=1}^{n}x_{j}=1$ and $\sum_{j=1}^{n}x_{j}^{2}=b^{2}$; then
\begin{equation}
\max\bigl\{x_{j}:{\>}1\leq{j}\leq{n}\bigr\}\leq\frac{1}{n}\left(1+\sqrt{n-1}\sqrt{n{b}^{2}-1}\right)
{\,}. \label{aust}
\end{equation}
\end{app2}

{\bf Proof.} For definiteness, we will suppose that
$x_{j}\leq{x}_{n}$ for all $j=1,\ldots,n-1$. Applying Jensen's
inequality to the convex function $x\mapsto{x}^{2}$, we write
\begin{equation}
\left(\frac{1}{n-1}\sum\nolimits_{j=1}^{n-1}x_{j}\right)^{2}\leq
\frac{1}{n-1}\sum\nolimits_{j=1}^{n-1}x_{j}^{2}
\ . \label{jens0}
\end{equation}
From one of the preconditions, we have
$\sum_{j=1}^{n-1}x_{j}=1-x_{n}$. Combining this with equation
(\ref{jens0}) finally leads to the inequality
\begin{equation}
\frac{(1-x_{n})^{2}}{n-1}\leq\sum\nolimits_{j=1}^{n-1}x_{j}^{2}
\ . \label{jens1}
\end{equation}
From another precondition, we further write the restriction
\begin{equation}
(1-x_{n})^{2}+(n-1)x_{n}^{2}=n\Bigl(x_{n}-\frac{1}{n}\Bigr)^{2}+\frac{n-1}{n}\leq(n-1)b^{2}
\ , \label{jens2}
\end{equation}
which can be expressed as
$$
\bigl(x_{n}-n^{-1}\bigr)^{2}\leq{n}^{-2}(n-1)\bigl(n{b}^{2}-1\bigr)
\ .
$$
Taking the square root, the number $x_{n}$ is bounded from above
by the right-hand side of equation (\ref{aust}).
$\blacksquare$

An upper bound on vector $\infty$-norms follows from equation (\ref{aust}).
For $q\geq1$, the $q$-norm of vector $u\in{\mathbb{C}}^{n}$ is
defined as
\begin{equation}
\|u\|_{q}:=\left(\sum\nolimits_{j=1}^{n}|u_{j}|^{q}\right)^{1/q}
. \label{qnrm}
\end{equation}
The case $q=2$ gives the usual Euclidean norm $\|u\|_{2}$. In the
limit $q\to\infty$, the definition (\ref{qnrm}) leads to the norm
\begin{equation}
\|u\|_{\infty}:=\max\bigl\{|u|_{j}:{\>}1\leq{j}\leq{n}\bigr\}
\ . \label{innrm}
\end{equation}
Substituting $x_{j}=\|u\|_{1}^{-1}|u_{j}|$ into equation
(\ref{aust}), we immediately obtain the inequality
\begin{equation}
\|u\|_{\infty}\leq\frac{1}{n}\left(\|u\|_{1}+\sqrt{n-1}\sqrt{n\|u\|_{2}^{2}-\|u\|_{1}^{2}}\right)
{\,}. \label{auust}
\end{equation}
The inequality (\ref{auust}) is clearly saturated in the following
two cases. In the first, only one of the vector components is
non-zero, whence all the $q$-norms coincide. In the second, the
vector components have the same absolute value $\|u\|_{1}/n$,
whence $\|u\|_{q}=n^{(1-q)/q}{\,}\|u\|_{1}$. Similar relations can
be written for the corresponding Schatten norms of a linear
operator on $\hh$. We refrain from presenting the details here.

\section{Details related to the condition (\ref{prcp})}\label{appr}

In this appendix, we prove that the transformation with elements
(\ref{qipj1}) obeys equation (\ref{prcp}) for arbitrary input
tuple $u$. First, we prove equation (\ref{prcp}) in the case of
pure state $\bro=|\psi\rangle\langle\psi|$. With given
$|\psi\rangle$, we can take phase factors in the vectors
$|m_{i}\rangle$ and $|n_{j}\rangle$ in such a way that the inner
products $\langle\psi|m_{i}\rangle$ and $\langle\psi|n_{j}\rangle$
are all real positive. Then the transformation elements become
$t_{ij}=\langle{m}_{i}|n_{j}\rangle$. Recall that any POVM with
rank-one elements can be realized via an orthogonal basis in the
space of corresponding dimension (see, e.g., sect. 3.1 of
reference \cite{preskill}). By $n_{ij}$, we denote the components
of $j$th vector $|n_{j}\rangle$. Due to the completeness relation,
$d$ rows of the $d\times{N}$-matrix $[[n_{ij}]]$ are mutually
orthogonal. By adding rows, this matrix can be converted into a
unitary $N\times{N}$-matrix. Its columns written as
$|n_{j}\rangle\oplus|n_{j}^{\perp}\rangle$ form an orthonormal
basis in $N$-dimensional space $\hh\oplus\hh_{\nc}$. Similarly, we
obtain an orthonormal basis of the vectors
$|m_{i}\rangle\oplus|m_{i}^{\perp}\rangle$ in $M$-dimensional
space $\hh\oplus\hh_{\mc}$, where $M=|\mc|$. We now define the two
sets of vectors, lying in the space
$\hh\oplus\hh_{\mc}\oplus\hh_{\nc}$, namely
\begin{equation}
|\widetilde{m}_{i}\rangle:=
\begin{pmatrix}
 |m_{i}\rangle \\
 |m_{i}^{\perp}\rangle \\
 \bnil
\end{pmatrix}
{\,}, \qquad
|\widetilde{n}_{j}\rangle:=
\begin{pmatrix}
 |n_{j}\rangle \\
 \bnil \\
 |n_{j}^{\perp}\rangle
\end{pmatrix}
{\,}. \label{wmn}
\end{equation}
Here, columns $\bnil$ denote zero vectors of the corresponding
dimensionality. The sets $\bigl\{|\widetilde{m}_{i}\rangle\bigr\}$
and $\bigl\{|\widetilde{n}_{j}\rangle\bigr\}$ are both orthonormal
and incomplete. By orthonormality, for any tuple $u$ we have
\begin{equation}
\sum\nolimits_{j}|u_{j}|^{2}=\langle\widetilde{w}|\widetilde{w}\rangle
\ , \qquad
|\widetilde{w}\rangle=\sum\nolimits_{j}u_{j}{\,}|\widetilde{n}_{j}\rangle
\ . \label{uwwu}
\end{equation}
Obviously, the unit vectors (\ref{wmn}) obey
$t_{ij}=\langle{m}_{i}|n_{j}\rangle=\langle\widetilde{m}_{i}|\widetilde{n}_{j}\rangle$.
Hence, the numbers
$v_{i}=\sum\nolimits_{j}t_{ij}{\,}u_{j}=\langle\widetilde{m}_{i}|\widetilde{w}\rangle$
are components of the orthogonal projection of
$|\widetilde{w}\rangle$ onto the subspace $\hh\oplus\hh_{\mc}$.
The squared modulus of this projection is equal to
$\sum\nolimits_{i}|v_{i}|^{2}$ and not larger than
$\langle\widetilde{w}|\widetilde{w}\rangle$.

A parallel approach works for mixed states. To given density
matrix
$\sum_{\lambda}\lambda{\,}|\psi_{\lambda}\rangle\langle\psi_{\lambda}|=\bro\in\lsp(\hh)$,
we assign the density matrix on the space
$\hh\oplus\hh_{\mc}\oplus\hh_{\nc}$:
\begin{equation}
\wbro:=\sum\nolimits_{\lambda}\lambda{\,}|\widetilde{\psi}_{\lambda}\rangle\langle\widetilde{\psi}_{\lambda}|
\ , \qquad
|\widetilde{\psi}_{\lambda}\rangle:=|\psi_{\lambda}\rangle\oplus\bnil\oplus\bnil
\ . \label{wrdf}
\end{equation}
Obviously, we have
$\langle{m}_{i}|\bro|n_{j}\rangle=\langle\widetilde{m}_{i}|\wbro|\widetilde{n}_{j}\rangle$
and similarly for other matrix elements of $\bro$. Further, we
introduce linear operators
\begin{align}
\wbom_{i}&:=\langle\widetilde{m}_{i}|\wbro|\widetilde{m}_{i}\rangle^{-1/2}{\,}
|\widetilde{m}_{i}\rangle\langle\widetilde{m}_{i}|{\,}\wbro^{1/2}
\ , \label{wots0}\\
\weta_{j}&:=\langle\widetilde{n}_{j}|\wbro|\widetilde{n}_{j}\rangle^{-1/2}{\,}
|\widetilde{n}_{j}\rangle\langle\widetilde{n}_{j}|{\,}\wbro^{1/2}
\ . \label{wots}
\end{align}
They clearly satisfy
$\langle\wbom_{i}{\,},\wbom_{i}\rangle_{\mathrm{hs}}=\langle\weta_{j}{\,},\weta_{j}\rangle_{\mathrm{hs}}=1$
and also
\begin{equation}
\langle\wbom_{i}{\,},\wbom_{k}\rangle_{\mathrm{hs}}=\delta_{ik}
\ , \qquad
\langle\weta_{j}{\,},\weta_{l}\rangle_{\mathrm{hs}}=\delta_{jl}
\ . \label{orhs}
\end{equation}
Here, we used orhonormality of
$\bigl\{|\widetilde{m}_{i}\rangle\bigr\}$ and
$\bigl\{|\widetilde{n}_{j}\rangle\bigr\}$. In terms of these
operators, the matrix elements are
\begin{equation}
t_{ij}=\frac{\langle\widetilde{m}_{i}|\widetilde{n}_{j}\rangle{\,}\langle\widetilde{n}_{j}|\wbro|\widetilde{m}_{i}\rangle}
{\langle\widetilde{m}_{i}|\wbro|\widetilde{m}_{i}\rangle^{1/2}{\,}\langle\widetilde{n}_{j}|\wbro|\widetilde{n}_{j}\rangle^{1/2}}
=\langle\wbom_{i}{\,},\weta_{j}\rangle_{\mathrm{hs}}
\ . \label{tijrw}
\end{equation}
For given input tuple $u$, the output numbers are expressed as
$v_{i}=\sum_{ij}t_{ij}{\,}u_{j}=\langle\wbom_{i}{\,},\wsig\rangle_{\mathrm{hs}}$,
where $\wsig=\sum_{j}u_{j}{\,}\weta_{j}$. Due to equation
(\ref{orhs}), this operator is also represented as
\begin{equation}
\wsig=\sum\nolimits_{i}v_{i}{\,}\wbom_{i}+\wvpi
\ , \label{wsit}
\end{equation}
where $\langle\wbom_{i}{\,},\wvpi\rangle_{\mathrm{hs}}=0$ for all
$i$. Calculating the squared Hilbert--Schmidt norms, we finally
obtain
\begin{equation}
\sum\nolimits_{i}|v_{i}|^{2}=
\Bigl\langle\sum\nolimits_{i}v_{i}{\,}\wbom_{i}{\,},\sum\nolimits_{i}v_{i}{\,}\wbom_{i}\Bigr\rangle_{\mathrm{hs}}
\leq\langle\wsig{\,},\wsig\rangle_{\mathrm{hs}}=\sum\nolimits_{j}|u_{j}|^{2}
\ . \label{finnm}
\end{equation}
Here, the orthonormality property (\ref{orhs}) is essential. Thus,
the precondition (\ref{prcp}) holds for all inputs.


\begin{thebibliography}{77}

\bibitem{heisenberg}
W.~Heisenberg, Z. Phys. {\bf 43}, 172 (1927)

\bibitem{lahti}
P.~Busch, T.~Heinonen, P.J.~Lahti, Phys. Rep. {\bf 452}, 155 (2007)

\bibitem{maass}
H.~Maassen, J.B.M.~Uffink, Phys. Rev. Lett. {\bf 60}, 1103 (1988)

\bibitem{ww10}
S.~Wehner, A.~Winter, New J. Phys. {\bf 12}, 025009 (2010)

\bibitem{brud11}
I.~Bia{\l}ynicki-Birula, {\L}.~Rudnicki, Entropic uncertainty
relations in quantum physics. In Sen, K.D. (ed.): Statistical
Complexity, 1--34. Springer, Berlin (2011)

\bibitem{shih03}
Y.~Shih, Eur. Phys. J. D {\bf 22}, 485 (2003)

\bibitem{dmg07}
I.~Damg{\aa}rd, S.~Fehr, R.~Renner, L.~Salvail, C.~Schaffner, A tight high-order entropic quantum uncertainty
relation with applications. In: Advances in Cryptology -- CRYPTO '07, Lecture Notes in Computer Science, vol. 4622, 360--378. Springer, Berlin (2007)

\bibitem{aamb10}
A.~Ambainis, Quantum Inf. Comput. {\bf 10}, 0848 (2010)

\bibitem{ngbw12}
H.Y.N.~Ng, M.~Berta, S.~Wehner, Phys. Rev. A {\bf 86}, 042315 (2012)

\bibitem{rprz12}
W.~Roga, Z.~Pucha{\l}a, {\L}.~Rudnicki, K.~\.{Z}yczkowski, Phys. Rev. A {\bf 87}, 032308 (2013)

\bibitem{molm09a}
S.~Wu, S.~Yu, K.~M{\o}lmer, Phys. Rev. A {\bf 79}, 022320 (2009)

\bibitem{petz10}
D.~Petz, J. Math. Phys. {\bf 51}, 015215 (2010)

\bibitem{schwinger}
J.~Schwinger, Proc. Natl. Acad. Sci. {\bf 46}, 570 (1960)

\bibitem{wf89}
W.K.~Wootters, B.D~Fields, Ann. Phys. {\bf 191}, 363 (1989)

\bibitem{gott96}
D.~Gottesman, Phys. Rev. A {\bf 54}, 1862 (1996)

\bibitem{cald97}
A.R.~Calderbank, E.M.~Rains, P.W.~Shor, N.J.A.~Sloane, Phys. Rev. Lett. {\bf 78} 405 (1997)

\bibitem{shbah12}
C.~Spengler, M.~Huber, S.~Brierley, T.~Adaktylos, B.C.~Hiesmayr, Phys. Rev. A {\bf 86}, 022311 (2012)

\bibitem{vaid87}
L.~Vaidman, Y.~Aharonov, D.Z.~Albert, Phys. Rev. Lett. {\bf 58}, 1385 (1987)

\bibitem{ena01}
B.-G.~Englert, Y.~Aharonov, Phys. Lett. A {\bf 284}, 1 (2001)

\bibitem{bz10}
T.~Durt, B.-G.~Englert, I.~Bengtsson, K.~\.{Z}yczkowski, Int. J. Quantum Inf. {\bf 8}, 535 (2010)

\bibitem{ivan92}
I.D.~Ivanovic, J. Phys. A: Math. Gen. {\bf 25}, L363 (1995)

\bibitem{sanchez93}
J.~S\'{a}nchez, Phys. Lett. A {\bf 173}, 233 (1993)

\bibitem{jsan95}
J.~S\'{a}nchez-Ruiz, Phys. Lett. A {\bf 201}, 125 (1995)

\bibitem{rastqip}
A.E.~Rastegin, Quantum Inf. Process. {\bf 12}, 2947 (2013)

\bibitem{ballester}
M.A.~Ballester, S.~Wehner, Phys. Rev. A {\bf 75}, 022319 (2007)

\bibitem{rbsc04}
J.M.~Renes, R.~Blume-Kohout, A.J.~Scott, C.M.~Caves, J. Math. Phys. {\bf 45}, 2171 (2004)

\bibitem{rottler}
A.~Klappenecker, M.~R\"{o}tteler, e-print arXiv:quant-ph/0502031 (2005)

\bibitem{rastqqt}
A.E.~Rastegin, e-print arXiv:1210.6742 [quant-ph] (2012)

\bibitem{watrous1}
J.~Watrous, {\it Theory of Quantum Information.} (University of Waterloo, Waterloo, 2011) \\ http://www.cs.uwaterloo.ca/$\sim$watrous/CS766/

\bibitem{bb84}
C.H.~Bennett, G.~Brassard, Quantum cryptography: public key
distribution and coin tossing. In: Proceedings of IEEE
International Conference on Computers, Systems, and Signal
Processing, Bangalore, India, 175--179. IEEE, New York (1984)

\bibitem{bruss98}
D.~Bru{\ss}, Phys. Rev. Lett. {\bf 81}, 3018 (1998)

\bibitem{bbrv02}
S.~Bandyopadhyay, P.O.~Boykin, V.~Roychowdhury, F.~Vatan, Algorithmica {\bf 34}, 512 (2002)

\bibitem{kr04}
A.~Klappenecker, M.~R\"{o}tteler, Constructions of mutually unbiased bases. In: Finite Fields and Applications, Lecture Notes
in Computer Science, vol. 2948, 137--144. Springer, Berlin (2004)

\bibitem{wb05}
P.~Wocjan, T.~Beth, Quantum Inf. Comput. {\bf 5}, 93 (2005)

\bibitem{bz07}
I.~Bengtsson, W.~Bruzda, {\AA}.~Ericsson, J.-{\AA}.~Larsson, W.~Tadej, K.~\.{Z}yczkowski, J. Math. Phys. {\bf 48}, 052106
(2007)

\bibitem{peresq}
A.~Peres, {\it Quantum Theory: Concepts and Methods.} (Kluwer, Dordrecht, 1993)

\bibitem{loveb11}
L.~Loveridge, P.~Bush, Eur. Phys. J. D {\bf 62}, 297 (2011)

\bibitem{prug77}
E.~Prugove\u{c}ki, Int. J. Theor. Phys. {\bf 16}, 321 (1977)

\bibitem{busch91}
P.~Busch, Int. J. Theor. Phys. {\bf 30}, 1217 (1991)

\bibitem{dps04}
G.M.~D'Ariano, P.~Perinotti, M.F.~Sacchi, J. Opt. B: Quantum Semiclass. Opt. {\bf 6}, S497 (2004)

\bibitem{adf07}
D.M.~Appleby, H.B.~Dang, C.A.~Fuchs, e-print arXiv:0707.2071 [quant-ph] (2007)

\bibitem{krsw05}
A.~Klappenecker, M.~R\"{o}tteler, I.~Shparlinski, A.~Winterhof, J. Math. Phys. {\bf 46}, 082104 (2005)

\bibitem{renyi61}
A.~R\'{e}nyi, On measures of entropy and information. In: Proceedings of 4th Berkeley symposium on mathematical
statistics and probability. Vol. I, 547--561. University of California Press, Berkeley (1961)

\bibitem{ja04}
P.~Jizba, T.~Arimitsu, Ann. Phys. {\bf 312}, 17 (2004)

\bibitem{molm09}
S.~Wu, S.~Yu, K.~M{\o}lmer, Phys. Rev. A {\bf 79}, 022104 (2009)

\bibitem{MWB10}
P.~Mandayam, S.~Wehner, N.~Balachandran, J. Math. Phys. {\bf 51}, 082201 (2010)

\bibitem{tsallis}
C.~Tsallis, J. Stat. Phys. {\bf 52}, 479 (1988)

\bibitem{HC67}
J.~Havrda, F.~Charv\'{a}t, Kybernetika {\bf 3}, 30 (1967)

\bibitem{sf06}
S.~Furuichi, J. Math. Phys. {\bf 47}, 023302 (2006)

\bibitem{rastkyb}
A.E.~Rastegin, Kybernetika {\bf 48}, 242 (2012)

\bibitem{rchtf12}
R.~Chaves, T.~Fritz, Phys. Rev. A {\bf 85}, 032113 (2012)

\bibitem{bengtsson}
I.~Bengtsson, K.~\.{Z}yczkowski, {\it Geometry of Quantum States: An
Introduction to Quantum Entanglement.} (Cambridge University Press, Cambridge, 2006)

\bibitem{rastctp}
A.E.~Rastegin, Commun. Theor. Phys. {\bf 58}, 819 (2012)

\bibitem{ht01}
P.~Harremo\"{e}s, F.~Tops{\o}e, IEEE Trans. Inf. Theory {\bf 47}, 2944 (2001)

\bibitem{hardy}
G.H.~Hardy, J.E.~Littlewood, G.~Polya, {\it Inequalities.} (Cambridge University Press, London, 1934)

\bibitem{BCCRR10}
M.~Berta, M.~Christandl, R.~Colbeck, J.M.~Renes, R.~Renner, Nature Phys. {\bf 6}, 659 (2010)

\bibitem{ccyz12}
P.J.~Coles, R.~Colbeck, L.~Yu, M.~Zwolak, Phys. Rev. Lett. {\bf 108}, 210405 (2012)

\bibitem{birula06}
I.~Bia{\l}ynicki-Birula, Phys. Rev. A {\bf 74}, 052101 (2006)

\bibitem{rast10r}
A.E.~Rastegin, J. Phys. A: Math. Theor. {\bf 43}, 155302 (2010)

\bibitem{rast12num}
A.E.~Rastegin, Quantum Inf. Comput. {\bf 12}, 0743 (2012)

\bibitem{rast104}
A.E.~Rastegin, J. Phys. A: Math. Theor. {\bf 44}, 095303 (2011)

\bibitem{rast12quasi}
A.E.~Rastegin, J. Phys. A: Math. Theor. {\bf 45}, 444026 (2012)

\bibitem{raja95}
A.K.~Rajagopal, Phys. Lett. A {\bf 205}, 32 (1995)

\bibitem{gparis06}
J.~Ghiglieri, M.G.A.~Paris, Eur. Phys. J. D {\bf 40}, 139 (2006)

\bibitem{massar07}
S.~Massar, Phys. Rev. A {\bf 76}, 042114 (2007)

\bibitem{krpr02}
M.~Krishna, K.R.~Parthasarathy, Sankhy\={a}, Ser. A {\bf 64}, 842 (2002)

\bibitem{preskill}
J.~Preskill, {\it Quantum Computation and Information.} (California
Institute of Technology, California, 1998)\\ http://www.theory.caltech.edu/people/preskill/ph229/


\end{thebibliography}
\end{document}